\documentclass[10pt]{article}%Defines the document class as article with a font size of 10 points.
\usepackage{fullpage} %Expands the text area to use nearly the full page by setting smaller margins.
\usepackage[section]{placeins}%Ensures that floats (figures, tables) do not move past their respective sections.
\usepackage{romannum}%use Roman numerals
\usepackage[superscript]{cite}%option modifies the appearance of citations

\usepackage[colorlinks = true,
linkcolor = blue,
urlcolor  = blue,
citecolor = blue,
anchorcolor = blue]{hyperref}

\usepackage{authblk}%Facilitates adding author affiliations and details in a structured way.
\usepackage{graphicx}%Provides commands for including graphics in the document.
\usepackage{amsmath}%Adds advanced mathematical symbols and font support.

\makeatletter
\renewcommand\@biblabel[1]{\textbf{#1.}\hfill}
\makeatother

\begin{document}
	
	\title{Double hysteresis loop in synchronization transitions of multiplex networks: the role of frequency arrangements and frustration}
	
	\author[1]{Ali Seif}
	\author[1,*]{Mina Zarei}
	\affil[1]{Institute of Advanced Studies in Basic Sciences (IASBS), Department of Physics, Zanjan, 45137-66731, Iran.}
	\vspace{-1em}

	\date{\today}%\date{November 17, 2024}

	\maketitle %Generates the title block using the current formatting. 
	\pagenumbering{arabic} %Switch to Arabic numerals
	
	\begin{abstract}
		\noindent%suppress the indentation of the first paragraph
		{This study explores the dynamics of two-layer multiplex networks, focusing on how frequency distributions among mirror nodes influence phase transitions and synchronization across layers. We present a Regular frequency assignment model for duplex networks, where the layers are fully connected and share identical sets of natural frequencies. By adjusting the sizes of sections where nodes exhibit positive, zero, or negative frequency differences relative to their mirrored equivalents in another layer, we can effectively control the average frequency discrepancy between the layers.
			We compared the dynamics of this structured model to those with randomly distributed frequencies, keeping a constant average frequency difference between the layers and introducing a phase lag in the interlayer interaction terms.  This comparison highlighted distinct behaviors, including double hysteresis loops in the synchronization phase transition and standing waves for intralayer coupling at the locations of the hysteresis loops, where the waves are composed of different interacting frequencies.
			\vspace{1em} % Add vertical space
			
			\noindent keywords: Explosive synchronization, Multiplex networks, Frustration, Phase-amplitude coupling}
	\end{abstract}

	\section*{Introduction}
	\phantomsection
	\label{Introduction}
	\addcontentsline{toc}{section}{Introduction}
	Multilayer networks have attracted considerable interest due to their prevalence in complex systems. Many real-world networks, including social, technological, and biological structures, can be represented as multilayer systems with different interactions across layers~\cite{shahal2020synchronization,bianconi2018multilayer,kivela2014multilayer,boccaletti2014structure,de2013mathematical,danziger2019dynamic}.  The brain serves as an excellent example of a multilayer network, with distinct layers representing various types of connectivity, such as structural, functional, and effective connections~\cite{vaiana2020multilayer,lim2019discordant,battiston2017multilayer}. Furthermore, the brain's multilayer framework encompasses a range of interactions between neurons, such as chemical and electrical synapses, along with volume transmission~\cite{bentley2016multilayer}. Different brain regions also communicate across diverse frequency bands, creating communication layers that facilitate various cognitive functions~\cite{buldu2018frequency,de2016mapping}. Thus, examining these multilayer structures can provide valuable insights into the dynamics and functionality of the brain.
	
	Synchronization is a fascinating phenomenon observed in various systems, including ecological, biological, and technological networks~\cite{strogatz2004sync,tang2014synchronization}. Researchers have studied the synchronization of networks made up of neurons and phase oscillators using diverse models~\cite{arenas2008synchronization,smeal2010phase,stiefel2016neurons,borgers2017introduction}. The Kuramoto model serves as a foundational framework for examining synchronization in oscillatory systems, providing insights into how oscillators synchronize based on their coupling strengths and natural frequencies~\cite{kuramoto1984chemical}. This model has been extended to incorporate realistic features of real-world systems, such as time delays~\cite{yeung1999time,ziaeemehr2020frequency}, inertia~\cite{filatrella2008analysis,mahdavi2024synchronization}, phase lags~\cite{sakaguchi1986soluble,sakaguchi1988mutual,mahdavi2022synchronization}, and interlayer interactions~\cite{nicosia2017collective,zhang2015explosive,jalan2019explosive,jalan2019inhibition,kumar2021explosive,kachhvah2019delay,kumar2020interlayer,jain2023composed,rathore2023synchronization,skardal2020higher,sadilek2015physiologically}.

	A key area of research centers on explosive synchronization, which involves a sudden shift to synchronized behavior. This phenomenon is crucial for numerous applications and has been linked to events such as cascading failures in power grids~\cite{buldyrev2010catastrophic}, coupled chemo-mechanical systems~\cite{kumar2015experimental}, and chronic pain or epileptic seizures in the brain~\cite{lee2018functional,adhikari2013localizing}.  Previous studies have explored the characteristics and structures that contribute to explosive synchronization in complex networks~\cite{boccaletti2016explosive,zhang2013explosive,d2019explosive}. For example, in single-layer networks, a correlation between node degree and natural frequency has been shown to promote explosive synchronization~\cite{gomez2011explosive,papadopoulos2017development}. However, this correlation is not necessary; explosive synchronization can also occur in adaptive networks that do not exhibit such a relationship~\cite{zhang2015explosive}. Additionally, abrupt synchronization transitions have been noted in fully connected networks with bimodal natural frequency distributions~\cite{martens2009exact}.  Furthermore, explosive synchronization has been observed in multiplex networks~\cite{nicosia2017collective,zhang2015explosive,jalan2019explosive,jalan2019inhibition,kumar2021explosive,kachhvah2019delay,khanra2018explosive,soriano2019explosive,wu2022double,jalan2020explosive,kachhvah2021explosive,kachhvah2020interlayer}.

	In this study, we explore how frequency arrangements in duplex networks influence their dynamics. We find that not only does the average frequency mismatch between mirror nodes impact the dynamics, but their specific configuration also plays a significant role. Previous research has demonstrated that explosive synchronization can arise from random frequency mismatches between mirror nodes across different layers, especially when interlayer interactions experience frustrations around $\frac{\pi}{2}$~\cite{kumar2021explosive}. Introducing frustrations or phase lags in the Kuramoto model's interaction terms typically alters the phase response curves of the oscillators, which subsequently affects the overall network dynamics~\cite{smeal2010phase,ziaeemehr2020emergence,skardal2015erosion}. Additionally, frustrations interact in complex ways with the natural frequencies of nodes during synchronization~\cite{brede2016frustration,xu2024maximal}.
	
	We propose a model of regular frequency arrangements that illustrates irreversibility and memory in the synchronization transition curves of duplex networks, even in the absence of phase lag in interlayer interactions. By introducing frustrations of approximately $\frac{\pi}{2}$ in the interlayer interaction terms, we observe transition curves featuring double hysteresis loops. We observe periodic synchrony within the area of intralayer couplings, where hysteresis loops emerge. The observed waves consist of multiple frequency bands that interact with each other. Specifically, the amplitude of faster waves modulates the phase of slower waves, illustrating a dependence between these frequency bands. We quantify the interaction between phase and amplitude through phase-amplitude coupling, a measure used for brain waves that demonstrates how different frequency bands interact across various time scales. This interaction facilitates complex cognitive processes and enhances efficient information processing.~\cite{seymour2017detection,daume2017phase,lega2016slow}.

	The paper is organized as follows: The first section offers a comprehensive overview of the structure and dynamical models, detailing the formulas and metrics utilized in our study. Section~\hyperref[results]{results} presents the numerical findings, emphasizing the synchronization transition curve of duplex networks, both with and without frustration in interlayer connections. We also investigate the periodic behavior noted in specific intralayer links and the interactions between the various oscillatory rhythms of the observed waves. Finally, the~\hyperref[Discussion]{discussion} section highlights our main conclusions.

	%##########################################################
	%########										   ########
	%########			MODEL AND METHOD	      	   ########
	%########										   ########
	%##########################################################
	\section*{Material and methods}
	\phantomsection
	\label{Material and Methods}
	\addcontentsline{toc}{section}{Material and Methods}
	\subsection*{Network topology and dynamics}
	\phantomsection
	\addcontentsline{toc}{subsection}{Network Topology and Dynamics}

\begin{figure}[!ht]
		\centering
		\includegraphics[width=0.44\linewidth]{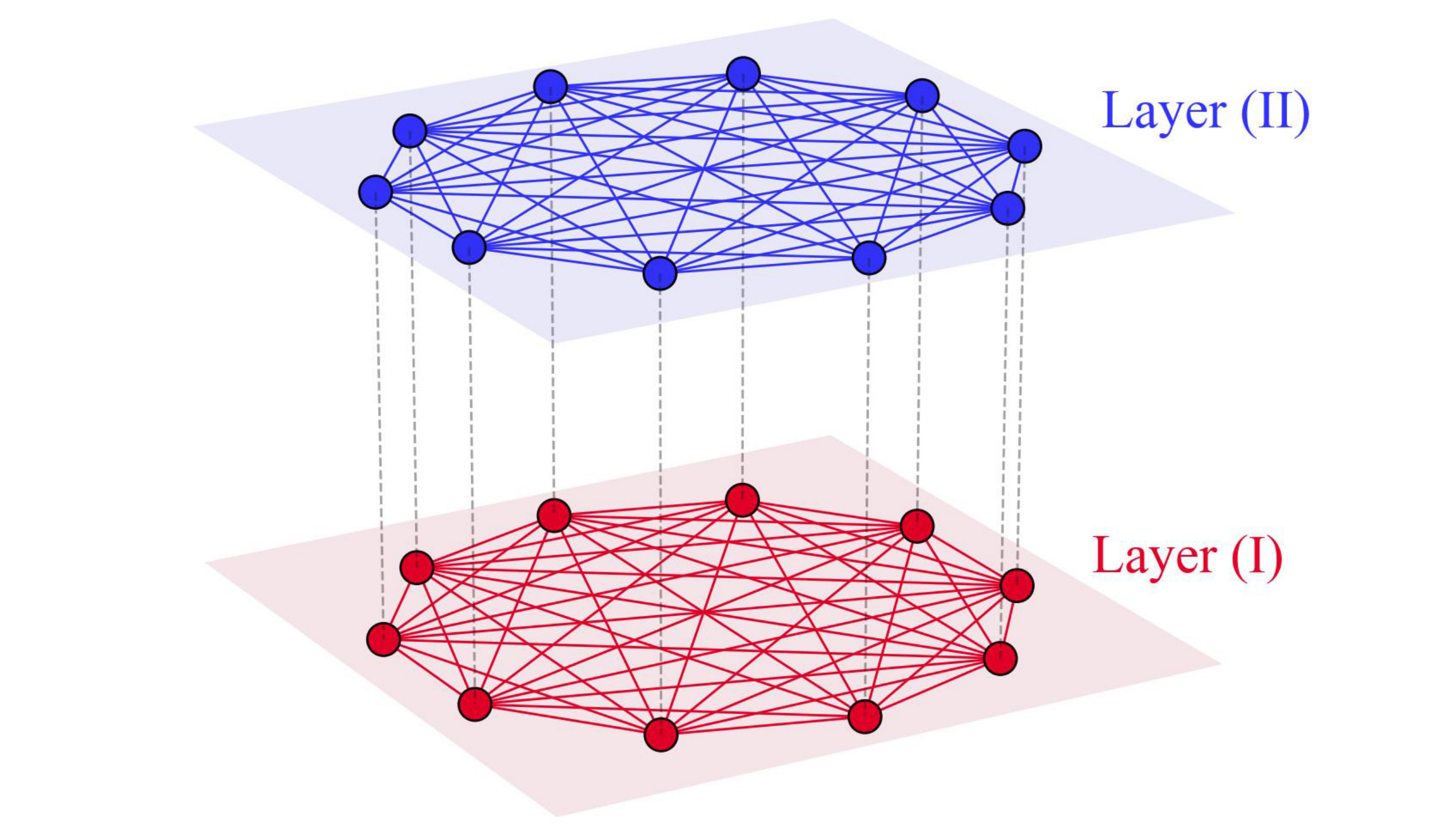}
		\caption{{Schematic representation of a duplex network with fully connected layers. Dashed lines indicate interlayer couplings between mirror nodes.}}
		\label{fig:fig1}
	\end{figure}

	Our implementation involves the utilization of a duplex network, a multiplex system consisting of two interconnected layers with all-to-all connectivity. We represent the nodes in each layer using Kuramoto oscillators as our chosen model. These oscillators serve as the nodes within each layer and capture their dynamic behavior. {\bf Mirror nodes} are pairs of nodes that represent the same entity across different layers of a multiplex network. In our model, these nodes are connected through an interlayer coupling mechanism that influences the synchronization and dynamics of the Kuramoto oscillators, enabling information exchange across the network’s layers. A visual representation of a duplex network can be found in \hyperref[fig:fig1]{Fig.~\ref*{fig:fig1}}. The interaction between mirror nodes across different layers is illustrated using dashed lines.
	
	The original Kuramoto model was developed for all-to-all interactions~\cite{kuramoto1984chemical}; however, it has been adapted for a variety of network structures over time~\cite{arenas2008synchronization}. In this study, we utilize an extended version of the Kuramoto model to investigate the dynamics of a duplex network comprising two interconnected layers ~\cite{kumar2021explosive}. The phase evolution of the $i$\textsuperscript{th} node in layer~\Romannum{1} and its mirror in layer~\Romannum{2} is described by the following equations:
\begin{subequations}
\begin{align}
	\label{eq:1-a}
	\dot{\theta_i^{I}} &=\omega_i^{I}+\frac{\sigma}{N} \sum_{j=1}^N \sin \left(\theta_j^{I}-\theta_i^{I}\right)+\lambda \sin \left(\theta_i^{II}-\theta_i^{I}+\alpha\right), \\
	\label{eq:1-b}
	\dot{\theta_i^{II}} &=\omega_i^{II}+\frac{\sigma}{N} \sum_{j=1}^N \sin \left(\theta_j^{II}-\theta_i^{II}\right)+\lambda \sin \left(\theta_i^{I}-\theta_i^{II}+\alpha\right),
\end{align}
\label{eq:1}
\end{subequations}\\
	where $\theta_i^{{I}({II})}$ indicates the phase of the $i$\textsuperscript{th} node in layer~\Romannum{1}(\Romannum{2}), and $\omega_i^{{I}({II})}$ represents the natural frequency of the $i$\textsuperscript{th} node in layer~\Romannum{1}(\Romannum{2}). The intralayer coupling strength, $\sigma$, is identical for both layers, and the interlayer coupling strength is denoted by $\lambda$.  The variable $N$ represents the total number of nodes in each layer of the duplex network. A frustration parameter, represented as $\alpha$, is incorporated into the interlayer interaction terms to account for phase lags between corresponding mirror nodes, thereby influencing their synchronization dynamics.
	
	To quantify the degree of synchronization within each layer, the synchronization order parameter is defined as follows:
	%\begin{center}
	\begin{equation}
		r^{{I}({II})} e^{i \psi^{{I}({II})}} = \frac{1}{N} \sum_{j=1}^N e^{i \theta_j^{{I}({II})}}.
		\label{eq:32}
	\end{equation}
	%\end{center}
	The synchronization order parameter, represented as $r^{{I}({II})}$, ranges from 0 to 1. A value of 1 reflects complete synchronization in the respective layer, where all nodes are in phase, whereas a value of 0 indicates total asynchrony. $\psi^{{I}({II})}$ represents the collective phase value, which can range between $0$ and $2\pi$. The long-term average of the synchronization order parameter $r^{{I}({II})}$, denoted as $R^{{I}({II})}$, provides insights into the overall level of synchronization in the stationary state within each layer. In our simulation, we compute the average $R^{{I}({II})}$ over the last $80\%$ of the simulation duration, which reflects the period when the system has reached its stationary state.
	
	To investigate synchrony within specific subsets of nodes in each layer, we define a local order parameter $R_{g}(t)$, where $g$ denotes different groups of oscillators. This parameter quantifies synchrony among  selected oscillators, providing insights into localized behaviors that may vary from the overall dynamics of the network. The calculation method is as follows:
	\begin{equation}
		R_{g}(t)=\frac{1}{N_{g}} \left|\sum_{j} e^{i\theta_{j}(t)}\right|,
		\label{eq:RP}
	\end{equation}
	here, $N_g$ represents the number of nodes in the selected group, and the sum is computed over these nodes.
	
	To assess the instantaneous correlation between the phases of all node pairs at a specific time instance, we define the correlation matrix $D(t)$ as follows:
	\begin{equation}
		D_{i j}(t)= \cos \left(\theta_i(t)-\theta_j(t)\right).
		\label{eq:D}
	\end{equation}\\
	The correlation matrix contains values ranging from $-1$ to $1$, where a value of $1$ signifies an in-phase relationship and a value of $-1$ indicates an anti-phase relationship between nodes $i$ and $j$ at the specified time point.
	
	In our simulation, we consider $N = 1000$. The initial phases of the oscillators are randomly sampled from a uniform distribution within the range $-\pi < \theta_{i}^{{I}({II})} \leq \pi$. To achieve the results, we numerically solve  \hyperref[eq:1]{Equations~(\ref*{eq:1})} employing the fourth-order Runge-Kutta method, with a time step of $dt=0.01$. Each simulation is run for a total of $40,000$ steps.
	
	\subsection*{Frequency arrangement models for mirror nodes in duplex networks}
	\phantomsection
	\addcontentsline{toc}{subsection}{Frequency Arrangement Models for Mirror Nodes in Duplex Networks}
	
	We have outlined the structure and dynamical model used in our study, along with methods for quantifying emergent behavior. Our main objective is to investigate how frequency arrangements among mirror nodes influence the collective behavior of the duplex network. To achieve this, we employ two different models: the {\bf Random model} and the {\bf Regular model}. To ensure comparability, the average natural frequency differences between mirror nodes in the two layers of both models can be maintained, even though frequency patterns may vary at a microscopic level. In the following, we will describe two models in detail and explain their differences.
	
	The natural frequencies in each layer are chosen as evenly spaced values ranging from $-0.5$ to $0.5$. In this paper, for both models, the natural frequencies of the nodes in layer~\Romannum{1} are organized in ascending order, with node $i=1$ assigned the minimum frequency of $-0.5$ and node $i=N$ assigned the maximum frequency of $0.5$. The following formula is used to establish these frequencies in the layer~\Romannum{1}:
	\begin{equation}
		\omega_{i}^{{I}} = -0.5 + \frac{i-1}{N-1}.
		\label{eq:2}
	\end{equation}

	The set of natural frequencies in the layer~\Romannum{2}  is identical to that in the layer~\Romannum{1}, differing only in the arrangement of these frequencies among the nodes within this layer. This variation in configuration can lead to differences in the frequencies of mirror nodes, meaning that $\omega_{i}^{I}$ may not be equal to $\omega_{i}^{I}$. As a result, there can be a non-zero average frequency difference between mirror nodes in the two layers. The average differences in natural frequencies between mirror nodes can be quantified as follows:
	\begin{equation}
		\Delta \omega=\frac{\sum_{i=1}^N\left|\delta\omega_i\right|}{2 \sum_{i=1}^N\left|\omega_i^{I}\right|},
		\label{eq:4}
	\end{equation}
	where $\delta\omega_i=\omega_i^{II}-\omega_i^{I}$; consequently, the value of $\Delta \omega$ ranges from 0 to 1. When mirror nodes have identical frequencies ($\omega_i^{I}= \omega_i^{II}$), $\Delta \omega$ is 0. In contrast, when mirror nodes have the maximum frequency difference ($\omega_i^{I}= -\omega_i^{II}$), $\Delta \omega$ is 1.
	As previously noted, the nodes in layer~\Romannum{1} are organized in ascending order according to their natural frequencies. Changes in the arrangement of node frequencies in layer~\Romannum{2} result in notable frequency discrepancies between the two layers, which can be quantified by the parameter $\Delta \omega$. The distinction between the Random and Regular models is based on the arrangement of frequencies in layer~\Romannum{2}, which is specifically configured to achieve a desired $\Delta \omega$.
	
	To implement the Random model for achieving a specific target value of $\Delta \omega$, the process begins by organizing the second layer in a sorted order, identical to that of the first layer, where $\Delta \omega = 0$. Following this initial arrangement, two nodes are randomly selected from the second layer, and their natural frequencies are swapped. If this swap results in a $\Delta \omega$ that is closer to the desired target, the change is accepted. Conversely, if the result does not bring the $\Delta \omega$ closer to the target, the frequencies are reverted to their previous configuration. This iterative adjustment continues until the intended $\Delta \omega$ value is successfully achieved.
	
	To achieve a targeted $\Delta \omega$ value using the Regular algorithm, we begin by organizing the nodes in both layers in ascending order based on their natural frequencies, initially setting $\Delta \omega$ to zero. In this study, we refer to pairs of nodes within each layer, labeled $i$ and $N+1-i$, as {\bf counterpart nodes}. In each step $S$, we select two counterpart nodes from layer~\Romannum{2} that demonstrate the greatest frequency differences with their corresponding mirror nodes in layer~\Romannum{1} and have not been previously selected, then swap their frequencies, specifically $\omega^{II}_{S}\longleftrightarrow\omega^{II}_{N+1-S}$. If the resultant $\Delta \omega$ is equal to or less than the desired value, we retain the change. Conversely, if the deviation exceeds the intended $\Delta \omega$, we revert the nodes to their original frequencies. This iterative process continues until we successfully achieve the specified $\Delta \omega$ value.

	\begin{figure}[!ht]
		\centering
		\includegraphics[width=0.86\linewidth]{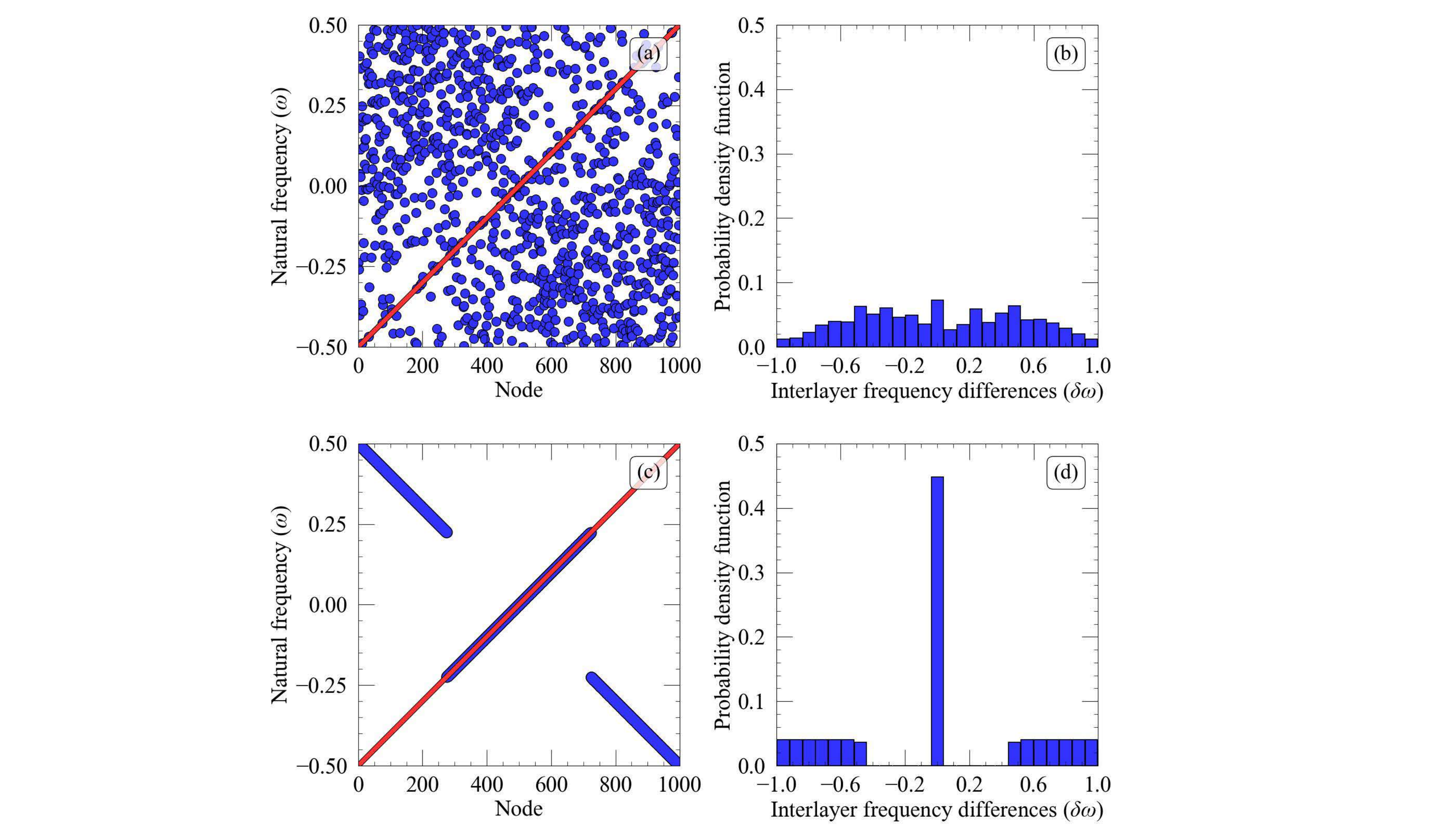}
		\caption{A visual representation showing the differences between Random and Regular models for frequency arrangement in a duplex network, both with $\Delta \omega=0.8$.
			(Left column) Red dots illustrate the natural frequencies of the nodes in layer~\Romannum{1}, which are arranged in ascending order. Blue dots illustrate the natural frequencies of the nodes in layer~\Romannum{2} for the ({\bf a})  Random model and ({\bf c})   Regular model. Panel~({\bf c}) identifies three distinct groups of nodes with positive, zero, and negative frequency differences ($\delta\omega_i$),  referred to as the left, middle, and right parts, respectively. 
			(Right column) The distribution function representing the magnitude of frequency differences between the mirror nodes, for the ({\bf b})   Random model and ({\bf d})  Regular model.}
		\label{fig:fig2}
	\end{figure}

	The visual representation in \hyperref[fig:fig2]{Fig.~\ref*{fig:fig2}} demonstrates how both Random and Regular algorithms are applied to generate distinct frequency arrangements while maintaining a constant $\Delta \omega$ value of 0.8. As previously mentioned, the network consists of two fully connected layers, forming a duplex structure. In layer~\Romannum{1}, nodes are arranged in ascending order of their frequencies, corresponding to the red-colored secondary diagonal of the squares in \hyperref[fig:fig2]{Fig.~\ref*{fig:fig2}a,c}. In the Random model, the frequencies of nodes in layer~\Romannum{2} are randomly assigned, leading to an almost uniform distribution of frequency differences between corresponding mirror nodes (\hyperref[fig:fig2]{Fig.~\ref*{fig:fig2}a,b}). In contrast, in the Regular model, we can observe the presence of three distinct sets of nodes, which we will refer to as {\bf left},  {\bf middle}, and {\bf right} hereafter. The nodes in the middle part exhibit zero frequency differences with their mirror nodes in layer~\Romannum{1} ($\delta\omega_i=\omega_i^{II}-\omega_i^{I}=0$), while the nodes in the left and right parts display positive and negative frequency differences, respectively (see \hyperref[fig:fig2]{Fig.~\ref*{fig:fig2}c}). This indicates a multimodal distribution of frequency discrepancies, as shown in \hyperref[fig:fig2]{Fig.~\ref*{fig:fig2}d}.

	\subsection*{Phase-Amplitude Coupling analysis}
	\phantomsection
	\addcontentsline{toc}{subsection}{Phase-Amplitude Coupling Analysis}
	Neural oscillations are synchronized rhythmic patterns of electrical activity generated by neurons in the brain. {\bf Phase-Amplitude Coupling (PAC)} is an important concept in neuroscience, employed to explore how the phase of low-frequency neural oscillations relates to the amplitude of high-frequency brain waves ~\cite{vanhatalo2004infraslow}. This establishes a framework for organizing and integrating neural information. It's fascinating how low-frequency oscillations can provide a timing mechanism for integrating the more detailed information carried by high-frequency oscillations. This coordination is essential for various cognitive processes, including attention, memory formation, and sensory perception~\cite{seymour2017detection,daume2017phase,lega2016slow}. Multiple methods have been developed to measure PAC~\cite{tort2010measuring,cohen2008oscillatory,combrisson2020tensorpac,hulsemann2019quantification}. In this paper, we will focus on two of these methods.
	
	The modulation index of the {\bf Phase-Locking Value (PLV)} serves as a key metric for evaluating the coordination between the phase of a {\bf Low-Frequency Oscillation (LFO)} and the amplitude envelope of a {\bf High-Frequency Oscillation (HFO)}~\cite{lachaux1999measuring}. To compute the PLV, the phase time series of the low-frequency oscillation ($\phi_{l}$) and the amplitude envelope of the high-frequency oscillation ($\phi_{A_{h}}$) are first extracted from the original signal using the Hilbert transform. Once the extraction is complete, the modulation index is analyzed to evaluate the degree of synchronization between the LFO and HFO, employing the following formula:
	\begin{equation}
		PLV = \frac{\left|\sum_{j=1}^{T} e^{i( \phi_{l}(j) -\phi_{A_{h}}(j) )}\right|}{T}.
	\end{equation}\\
	The summation over all $T$ time points aggregates the complex values corresponding to the phase differences at each time point. The normalized absolute value of the summed complex number gives a number between 0 and 1. A PLV of 1 indicates perfect phase synchronization, meaning that the phase relationship between two signals is consistent across all time points. On the other hand, a PLV of 0 suggests no consistent phase relationship between the signals. Essentially, high PLV values indicate a correlation between the phase of low-frequency oscillations (LFOs) and the amplitude of high-frequency oscillations (HFOs), while low values suggest a lack of correlation.
	
	Visualization of correlation can be achieved by plotting unit-length vectors, $e^{i( \phi_{l}(j) -\phi_{A_{h}}(j))}$ on a polar plane at various time points. Each vector's angle represents the phase difference between the signals. The degree of correlation is represented by averaging these unit vectors. The magnitude of the resulting vector indicates the PLV. Its direction reflects the phase of the low-frequency signal at which the amplitude of the high-frequency signal is amplified.
	
	The {\bf Mean-Vector Length (MVL) } modulation index serves as an secondary metric for assessing the synchronization between the phase of LFOs and the amplitude of HFOs~\cite{canolty2006high}. To calculate this index, the phases of the LFO (\(\phi_{l}\)) and the amplitude envelope of the HFO (\(A_{h}\)) are obtained from frequency-filtered signals using the Hilbert transform. Subsequently, the phase angles are grouped into bins labeled by the index $j$, and the average amplitude of the HFO within each phase bin of the LFO, denoted as \(\left< A_{h} \right>_{\phi_{l}(j)}\), is computed and normalized according to the following formula:
	\begin{equation}
		P_{(j)} = \frac{\left< A_{h} \right>_{\phi_{l}(j)}}{ \sum_{k=1}^{B} \left< A_{h} \right>_{\phi_{l}(k)}},
	\end{equation}\\
	where \(B\) represents the total count of phase bins. The Shannon entropy, obtained from the histogram mentioned previously as, $ H(p) = -\sum_{j=1}^{B} P_{(j)} \log P_{(j)}$, provides valuable insights into the correlation between phase and amplitude. This information contributes to the calculation of the modulation index of MVL through the following formula:
	\begin{equation}
		MVL = 1 - \frac{H(p)}{\log B}.
	\end{equation}\\
	The modulation index varies between $0$ and $1$, with higher values signaling more pronounced phase-amplitude coupling. The graph of $P(j)$ against $j$ also visually illustrates the relationship between phase and amplitude.
	
	Through the calculation of PLV or MVL and plotting their corresponding results, one can determine the presence of phase-amplitude coupling between different frequency bands of a signal.

	\section*{Results}
	\phantomsection
	\label{results}
	\addcontentsline{toc}{section}{Results}
	
\begin{figure}[!ht]
		\centering
		\includegraphics[width=0.86\linewidth]{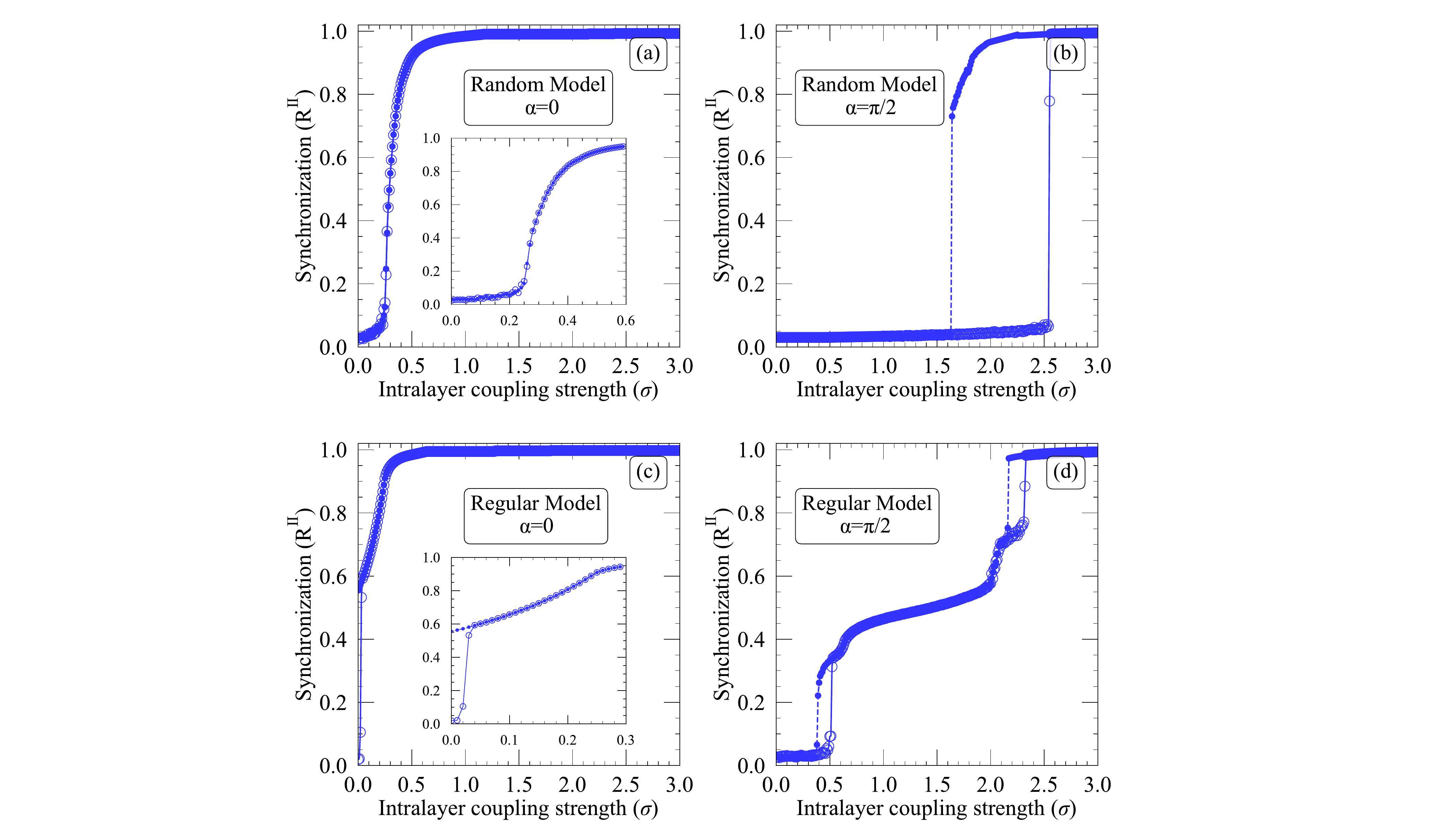}
		\caption{{Phase transitions from incoherent to coherent states occur as the intralayer coupling strength varies in a duplex network, utilizing two different frequency arrangement models. The results in the top row are associated with the Random model, while those in the bottom row pertain to the Regular model. The left column corresponds to $\alpha=0$, and the right column corresponds to $\alpha=\frac{\pi}{2}$. Solid lines with unfilled points indicate forward transitions, while dashed lines with filled points represent backward transitions. The parameters used are $\Delta \omega=0.8$ and $\lambda=10$. }}
		\label{fig:fig3}
	\end{figure}

	Our primary goal is to explore how patterns of frequency arrangements, which create frequency discrepancies between the two layers, influence the dynamics of a duplex network with frustrated interlayer interactions. \hyperref[fig:fig3]{Fig.~\ref*{fig:fig3}} depicts the transition from incoherent to coherent states in layer~\Romannum{2} of a duplex network, driven by intralayer coupling. The first row illustrates the phase transition for the Random model, whereas the second row presents the results for the Regular model. It's worth noting that the phase transition curves for both layers are quite similar, but we only present the results for the second layer here.

	In the Random model, the phase transition at $\alpha=0$ is continuous, while at $\alpha=\frac{\pi}{2}$, it appears discontinuous, exhibiting a hysteresis loop. This behavior aligns with the earlier findings of Anil Kumar et al., who demonstrated that the interaction between frequency discrepancies in the two layers and the frustration parameter near $\frac{\pi}{2}$ can lead to explosive synchronization~\cite{kumar2021explosive}.
	
	In the Regular model with $\alpha=0$, the transition curve is predominantly reversible across most intralayer couplings. However, at very low coupling values, a narrow hysteresis loop appears, indicating some irreversibility associated with the synchronization of the left and right sections. Beyond this range, the middle part synchronizes reversibly (see supplementary Fig.~SF1 and Fig.~SF2). This phenomenon likely results from the multimodal distribution of frequency discrepancies. In the backward curve, the nodes in both the right and left sections remain synchronized with zero frequency, even at $\sigma=0$. The supplementary video, SV1, illustrates the time evolution of the order parameter for the left, right, and middle sections, along with the correlation matrix at  $\sigma=0$. It clearly shows that the nodes in the left and right sections are synchronized at the average frequency, while the nodes in the middle section are not phase-locked.
	
	When a frustration of  $\alpha=\frac{\pi}{2}$ is applied to the Regular model, with $\Delta \omega=0.8$, a double hysteresis loop emerges in the phase transition curve.  This indicates that the transition from incoherent to coherent states occurs in two stages, each defined by distinct transition couplings, $\sigma^1_T$ and $\sigma^2_T$. The first transition takes place at a low intralayer coupling of  $\sigma^1_T=0.5$, and the second occurs at  $\sigma^2_T=2.3$ along the forward curve. \hyperref[fig:fig3]{Fig.~\ref*{fig:fig3}b} demonstrates that the Random model exhibits a single bistable region between incoherence and coherence at high coupling strengths. In contrast, the Regular model displays a bistable regime between asynchrony and partial synchronization at low coupling strengths, along with another regime transitioning from partial to complete synchronization at high coupling strengths. While the Random model remains asynchronous at low intralayer couplings, the Regular model can achieve partial synchronization. It's important to note that in both the Random and Regular models, all parameters remain constant, including  $\Delta \omega = 0.8$, which quantifies the average frequency discrepancy between mirror nodes in the two layers. The key distinction that explains the observed variations in transition curves lies in the specific patterns of frequency arrangements that create discrepancies between the layers.

	An analysis of the transition curves for the middle, left, and right sections reveals that the first observed hysteresis loop in \hyperref[fig:fig3]{Fig.~\ref*{fig:fig3}d} corresponds to the synchronization of the middle section, while the second loop is associated with the synchronization of the left and right sections, as illustrated in Supplementary Fig.~SF3. This result contrasts with the case of $\alpha=0$ , where the left and right sections synchronize at low intralayer couplings, while the middle section only starts to synchronize at higher couplings. In other words, positioning the frustration close to $\frac{\pi}{2}$ causes the arrangement of coupling configurations that synchronize the three different parts—left, middle, and right—to change along the transition curves (see supplementary Fig.~SF1 and Fig.~SF3 for comparison).
	
	As previously mentioned, earlier studies have shown that frequency discrepancies between layers in duplex networks with frustrated interlayer interactions can lead to explosive synchronization \cite{kumar2021explosive}. However, in the Regular model, the middle section—where the frequency discrepancy between mirror nodes is zero—also exhibits explosive synchronization (see \hyperref[fig:fig3]{Fig.~\ref*{fig:fig3}d}). This is probably due to the complex interaction between the multimodal distribution of frequency discrepancies and frustration.

\begin{figure}[!ht]
		\centering
		\includegraphics[width=0.86\linewidth]{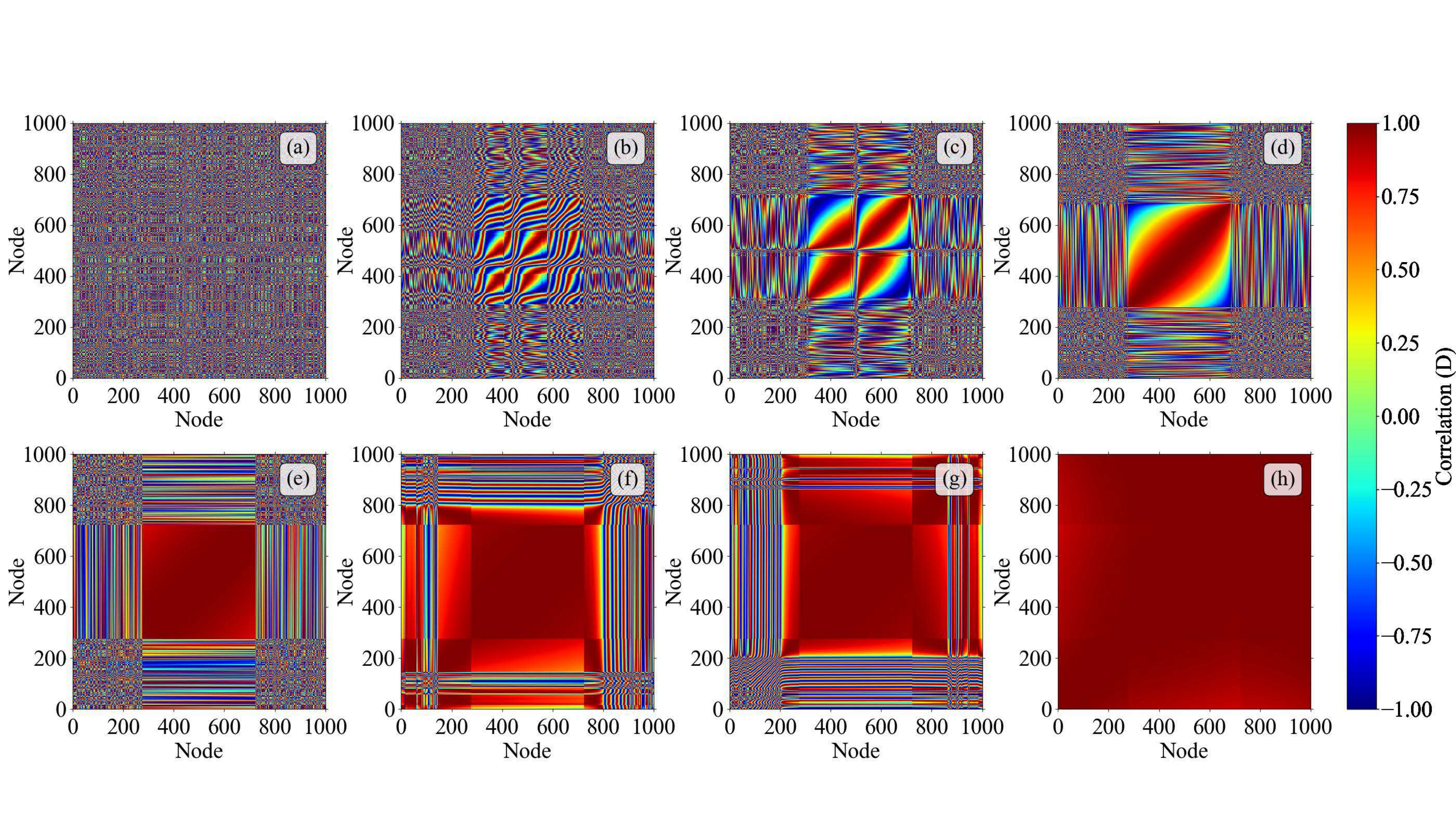}
		\caption{The instantaneous correlation matrices for layer~\Romannum{2} at different intralayer coupling strengths in the backward paths are presented. The figures are arranged from top left to bottom right, corresponding to intralayer coupling strengths ($\sigma$) of 2.71, 2.14, 2.13, 1.61, 0.60, 0.39, 0.38, and 0.09, respectively. This analysis used the Regular model for frequency  assignment and the same parameters as those in \hyperref[fig:fig3]{Fig.~\ref*{fig:fig3}d}.}
		\label{fig:fig4}
	\end{figure}

	To gain a comprehensive understanding of the transition depicted in \hyperref[fig:fig3]{Fig.~\ref*{fig:fig3}d}, we have plotted the instantaneous correlation matrix, $D$, for various values of the intralayer couplings in \hyperref[fig:fig4]{Fig.~\ref*{fig:fig4}}, focusing on the backward path in layer~\Romannum{2}.
	In fact, since the level of synchronization is higher on the backward path, we will have a better view of the details. In \hyperref[fig:fig4]{Fig.~\ref*{fig:fig4}}, red points indicate in-phase synchronization between the two nodes ($D_{i j}(t)=1$), while blue points represent anti-phase synchronization ($D_{i j}(t)=-1$).

	\hyperref[fig:fig4]{Fig.~\ref*{fig:fig4}} illustrate the changes in the correlation matrix patterns as the intralayer couplings are decreased, highlighting the transition from coherent to incoherent states. \hyperref[fig:fig4]{Fig.~\ref*{fig:fig4}a} shows that when there is strong intralayer coupling, such as $\sigma=2.71$, all nodes in layer~\Romannum{2} are nearly synchronized. \hyperref[fig:fig4]{Fig.~\ref*{fig:fig4}b,c} demonstrate that as the coupling is reduced to the region of the second hysteresis loop, the network exits the synchronized state; either the left part (\hyperref[fig:fig4]{Fig.~\ref*{fig:fig4}b}) or the right part (\hyperref[fig:fig4]{Fig.~\ref*{fig:fig4}c}) loses coherence, while the remaining parts stay synchronized. Then, as the intralayer coupling strength decreases further, for example to $\sigma=1.61$, only the nodes in the middle part—those have the same natural frequencies as their mirror nodes in the other layer—continue to synchronize (\hyperref[fig:fig4]{Fig.~\ref*{fig:fig4}d}). 
	As the coupling strength is reduced even more, the middle part also transitions out of the synchronized state (\hyperref[fig:fig4]{Fig.~\ref*{fig:fig4}e}). With the continuation of the declining process, in the first hysteresis loop, the nodes in the middle part create synchronized clusters of those with similar natural frequencies (\hyperref[fig:fig4]{Fig.~\ref*{fig:fig4}f,g} ). The less intralayer coupling there is, the more clusters there will emerge. Ultimately, with minimal intralayer coupling, we observe an incoherent state (\hyperref[fig:fig4]{Fig.~\ref*{fig:fig4}h}). In summary, the figure demonstrates that the first hysteresis loop is associated with the transition of the middle component to a synchronized state. In contrast, the second hysteresis loop corresponds to the synchronization of the left and right parts with the middle section. This finding aligns with the result shown in supplementary Fig.~SF3. It’s important to note that the correlation matrices shown are plotted at a single time point within stationary states. The question that arises is: how do the dynamics behave during stationary states? In continuation, we will concentrate on the dynamics within the region of the hysteresis loops.

\begin{figure}[!ht]
		\centering
		\includegraphics[width=0.86\linewidth]{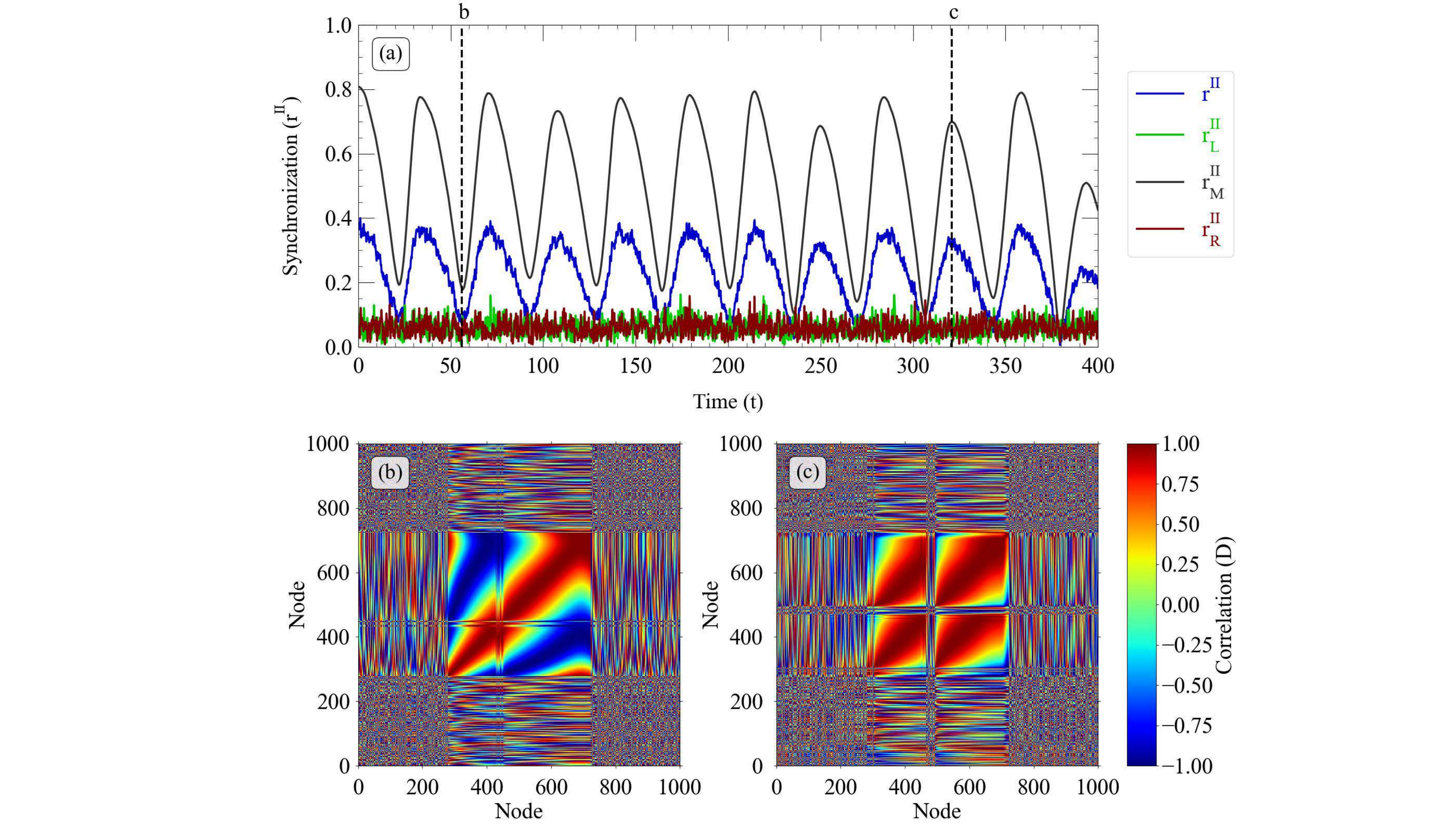}
		\caption{{A robust cyclic behavior in the order parameter is evident during the first hysteresis loop of the phase transition curve. ({\bf a}) The evolution of the synchronization order parameter in the Regular frequency assignment model is illustrated by blue, green, black, and red curves, representing the entire layer and its left, middle, and right parts, respectively. ({\bf b} and {\bf c}) Two snapshots of the correlation matrix at the time instances indicated by the dashed lines in the top panel. The parameters are identical to those shown in  \hyperref[fig:fig3]{Fig.~\ref*{fig:fig3}d}, with $\sigma=0.39$. }}
		\label{fig:fig5}
	\end{figure}

	\hyperref[fig:fig5]{Fig.~\ref*{fig:fig5}a} depicts the time evolution of the synchronization order parameter for the left, middle, and right parts of the network, as well as for the entire network, with the intralayer coupling set to $\sigma=0.39$ within the first hysteresis loop. It can be observed that the left and right parts are not synchronized. Moreover, the entire network displays oscillatory behavior, which is driven by the dynamics of the middle part. This oscillatory behavior is due to the fact that the middle part is divided into two nearly synchronized clusters with different frequencies. \hyperref[fig:fig5]{Fig.~\ref*{fig:fig5}b,c} present the correlation matrix at two instances corresponding to the peak and trough of the fluctuations, as indicated by dashed lines in \hyperref[fig:fig5]{Fig.~\ref*{fig:fig5}a}. During the peak of fluctuations, the two clusters in the middle region come close to each other and almost synchronize. In contrast, at the trough, the clusters are spaced apart, with all nodes evenly distributed around the unit circle according to their frequencies. For a detailed look at the dynamics, the time evolutions of the order parameter and correlation matrix can be viewed in supplementary video, SV2.

\begin{figure}[!ht]
		\centering
		\includegraphics[width=0.86\linewidth]{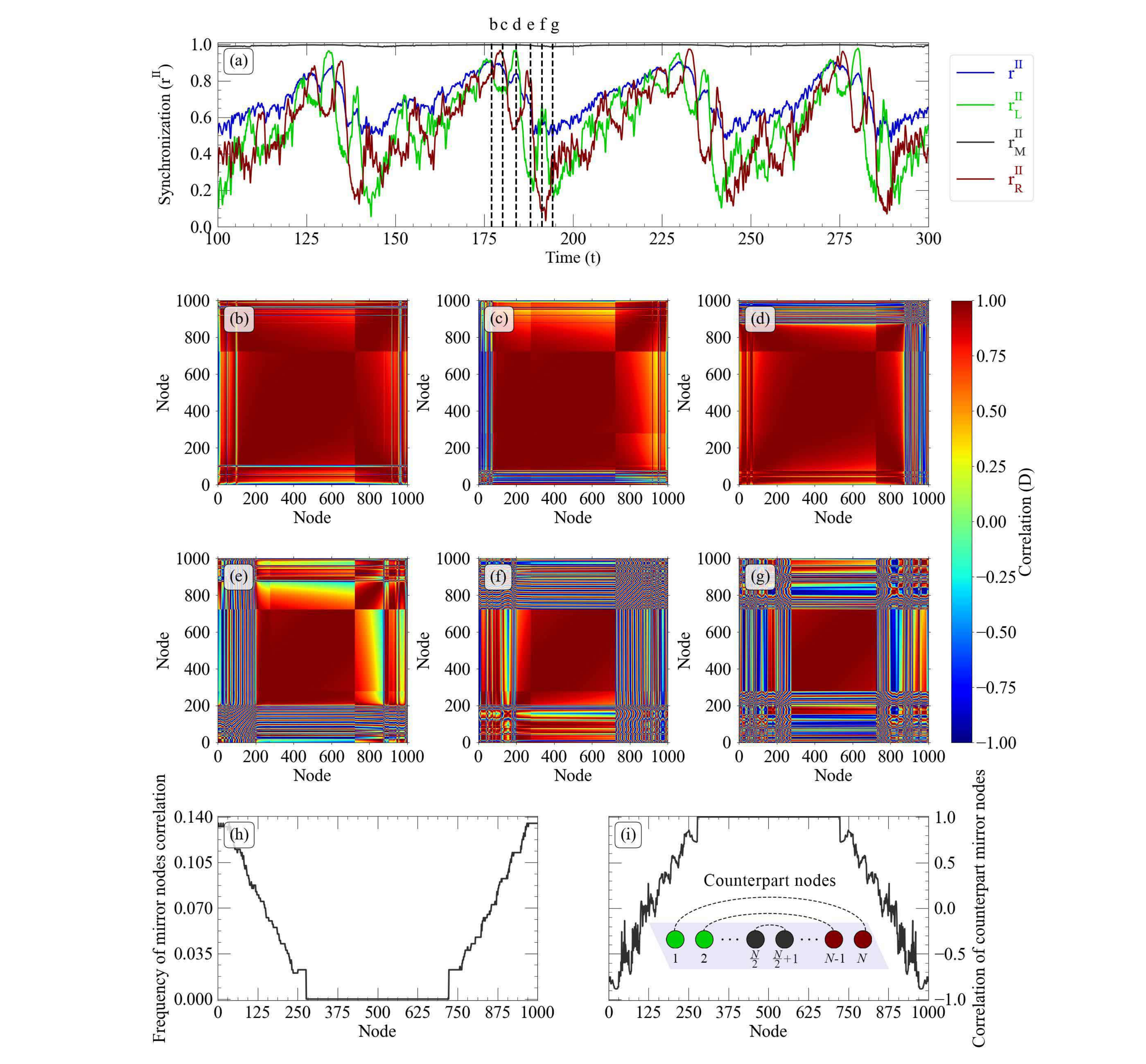}
		\caption{{Analyzing the intricate dynamics of each section of layer~\Romannum{2} in the second hysteresis loop of the phase transition curve. ({\bf a}) Observing the evolution of the synchronization order parameter for the entire layer, as well as for the left, middle, and right parts, represented by the blue, green, black, and red lines, respectively. ({\bf b}-{\bf g}) Six snapshots of the correlation matrix for the time points shown by the dashed lines in the top panel. ({\bf h}) Frequencies of phase correlations among mirror nodes. ({\bf i}) Phase correlation of the counterparts' mirror nodes. The same parameters as those used in \hyperref[fig:fig3]{Fig.~\ref*{fig:fig3}d} were applied, utilizing a value of $\sigma=2.11$.}}
		\label{fig:fig6}
	\end{figure}

	To analyze the dynamics of the network in the second hysteresis loop, we plot the time evolution of the synchronization order parameter for the left, middle, and right parts (green, black, and red lines, respectively), along with the entire layer~\Romannum{2} (blue line) at an intralayer coupling of $\sigma=2.11$ (see \hyperref[fig:fig6]{ Fig.~\ref*{fig:fig6}a}). We observe that while the middle part is fully synchronized, the synchronization of the entire layer exhibits periodic behavior influenced by the dynamics of the left and right sections.

	The dynamics of the left and right sections clearly show that different waves with various frequencies are superimposed. Notably, low-frequency oscillations~(LFOs) in the left and right parts are in-phase, while medium-frequency oscillations~(MFOs) are in anti-phase, exhibiting a phase difference of $\pi$ radians (see supplementary Fig.~SF4 for better illustration). To gain clearer insight into these dynamics, we have created plots of the instantaneous correlation matrix at different time points (highlighted by dashed lines in \hyperref[fig:fig6]{Fig.~\ref*{fig:fig6}a}), from one peak to the next trough of the overall order parameter in the second and third rows. As previously mentioned, the middle part is synchronized for intralayer coupling beyond the first transition point, and we are now focusing on the second hysteresis loop. In the second and third rows, a red square in the center of each panel clearly indicates the synchronized middle part. Now, we will focus on the corners of the correlation matrices related to the dynamics of the left and right parts.

	\hyperref[fig:fig6]{Fig.~\ref*{fig:fig6}b,g} display the correlation matrix at the time points when the entire layer (shown by the blue curve in \hyperref[fig:fig6]{Fig.~\ref*{fig:fig6}a}) reaches its maximum and minimum oscillation. This illustrates the synchronization of the left and right parts with the middle part, as shown in \hyperref[fig:fig6]{Fig.~\ref*{fig:fig6}b}, indicating almost full synchronization ($r\approx0.9$). In contrast, \hyperref[fig:fig6]{Fig.~\ref*{fig:fig6}g} shows the situation where the nodes on both sides are out of phase. 
	Thus, the complete synchronization and the out-of-phase behavior of the corner parts correspond to the peaks and troughs associated with the LFOs observed in the waves of the entire layer, as well as in the left and right regions.

	The correlation matrices in \hyperref[fig:fig6]{Fig.~\ref*{fig:fig6}c-f} depict the fluctuations between coherence and incoherence in the dynamics of the left and right parts during each LFO period. These panels illustrate how MFOs in the dynamics of the left and right parts—whose peaks are marked by dashed lines (c-f) in \hyperref[fig:fig6]{Fig.~\ref*{fig:fig6}a}—are superimposed on the low-frequency waves. Since the MFOs in the left and right parts are out of phase and effectively cancel each other out, they have no significant impact on the overall synchronization of the layer, as evidenced by the blue curve in \hyperref[fig:fig6]{Fig.~\ref*{fig:fig6}a} (refer to supplementary Fig. SF4 for a clearer illustration).

	\hyperref[fig:fig6]{Fig.~\ref*{fig:fig6}c,e} display correlation matrices for the peaks of MFOs in the right part's dynamics (red curve in \hyperref[fig:fig6]{Fig.~\ref*{fig:fig6}a}), which correspond to the troughs in the left part (green curve). In contrast, \hyperref[fig:fig6]{Fig.~\ref*{fig:fig6}d,f} present correlation matrices for the peaks (troughs) of MFOs in the left (right) part.

	By analyzing the process through the backward curve shown in \hyperref[fig:fig6]{Fig.~\ref*{fig:fig6}b-g}, we can observe the following dynamics: initially, the entire layer is synchronized (see \hyperref[fig:fig6]{ Fig.~\ref*{fig:fig6}b}). While the middle section remains synchronized, the outer parts of the left and right sections, which exhibit greater frequency differences between mirror nodes, unsynchronize and synchronize sequentially (\hyperref[fig:fig6]{Fig.~\ref*{fig:fig6}c,d} ). This blinking process spreads throughout both sections, leading to fluctuations in their dynamics (\hyperref[fig:fig6]{Fig.~\ref*{fig:fig6}e,f}). Finally, during the trough of the overall wave, both sections are simultaneously out of phase (\hyperref[fig:fig6]{Fig.~\ref*{fig:fig6}g}). This procedure illustrates the dynamics of transitioning from a peak to a trough in the left or right parts. As it moves from the trough to the next peak, the process appears to reverse, shifting from \hyperref[fig:fig6]{Fig.~\ref*{fig:fig6}g} to \hyperref[fig:fig6]{Fig.~\ref*{fig:fig6}b}, and this pattern repeats in the following cycles (see the supplementary video.~SV3 for a clear view of the blinking process).

	As mentioned earlier, each layer forms a fully connected network of Kuramoto oscillators, with the same sets of natural frequencies. In the absence of the duplex arrangement of layers with frequency discrepancies and phase frustration ($\alpha = \frac{\pi}{2}$) in interlayer interactions, we would expect a continuous phase transition between incoherent and synchronized states within each layer. The distinctive phase transition observed, characterized by double hysteresis loops and intricate periodic behavior in each layer, directly results from the network's duplex architecture, especially the Regular frequency assignment model for the mirror nodes.

	In \hyperref[fig:fig6]{Fig.~\ref*{fig:fig6}h,i}, we examine how the interaction between the two layers contributes to this complex dynamics. We achieve this by calculating the changes in phase correlations over time for each pair of mirror nodes, using the formula $C_i(t)=\cos(\theta_i^{I}(t)-\theta_i^{II}(t))$, for $i=1\dots N$, specifically focusing on intralayer coupling within the second hysteresis loop.  Our results show that the mirror nodes in the middle part  synchronize completely, while those on the left and right sides show periodic patterns.  \hyperref[fig:fig6]{Fig.~\ref*{fig:fig6}h} illustrates this periodicity by plotting the oscillation frequency of the correlation time series ($C_i(t)$) against the node number.The figure demonstrates that as the absolute value of frequency discrepancies between mirror nodes increases, the oscillation frequency in their correlation time series also rises. Thus, the periodic behavior observed in the evolution of the order parameter for both the left and right sections in each layer seems to be linked to the interactions between the layers, especially the complex dynamics of the mirror nodes within those sections.

	As shown in \hyperref[fig:fig3]{Fig.~\ref*{fig:fig3}c}, the frequency differences between the counterpart mirror nodes in the left and right parts exhibit similar magnitudes but opposite signs ($\delta\omega_{i}=-\delta\omega_{(N+1-i)}$, for $i=1,\dots,N$). This observation prompts an investigation into the potential relationship between the dynamics of the mirror nodes in the left and right sections. To explore this, we calculated the phase shift  between the correlation time series of the counterpart mirror nodes (i.e., $C_{i}(t)$ and $C_{N+1-i}(t)$) and plotted the cosine of this phase shift against the node number in \hyperref[fig:fig6]{Fig.~\ref*{fig:fig6}i}. The phase of the time series $C_{i}(t)$, denoted as $\phi_{i}(t)$, is determined using the Hilbert transform by calculating the arctangent of the analytic signal.  Since $\cos(\phi_{i}(t) - \phi_{N+1-i}(t)) = \cos(\phi_{N+1-i}(t) - \phi_{i}(t))$, this expression highlights the symmetry in the dynamics between the mirror pairs $i$ and $N+1-i$. Consequently, a symmetric plot is observed around the center at $i = \frac{N}{2} = 500$ in \hyperref[fig:fig3]{Fig.~\ref*{fig:fig6}i}.

	Each node in the middle part has a counterpart within the same part, whereas the counterparts of the nodes in the right part are found in the left part, and vice versa. As indicated in \hyperref[fig:fig6]{Fig.~\ref*{fig:fig6}h}, the nodes in the middle part are synchronized with each other and with their corresponding mirror nodes in the adjacent layer. Therefore, the counterpart mirror nodes associated with this part are correlated, as demonstrated by the values of $1$ shown in \hyperref[fig:fig6]{Fig.~\ref*{fig:fig6}i}. In other words, for the intralayer coupling within the second hysteresis loop we are analyzing in \hyperref[fig:fig6]{Fig.~\ref*{fig:fig6}}, there is no information transfer occurring among the nodes in the middle sections of each layer, nor between those sections across different layers.

	As shown in \hyperref[fig:fig6]{Fig.~\ref*{fig:fig6}h}, the mirror nodes in the left and right parts are not phase-locked, exhibiting periodic correlations with an oscillation period related to the magnitude of their natural frequency differences. Furthermore, each pair of mirror nodes in the left (right) part has a counterpart pair in the right (left) part that shares the same magnitude of natural frequency differences but with opposite signs. 
	Therefore, the correlation between the mirror nodes of both counterparts oscillates at similar frequencies, as illustrated in \hyperref[fig:fig6]{Fig.~\ref*{fig:fig6}h}; however, these oscillations are not in-phase.
	
	\hyperref[fig:fig6]{Fig.~\ref*{fig:fig6}i} illustrates that the correlation time series of the two counterpart pairs are shifted according to the magnitude of their natural interlayer frequency differences. Indeed, for the counterpart node pairs with the greatest frequency differences from their corresponding mirror nodes, they are in anti-phase with each other, as indicated by the values of ${-1}$ in \hyperref[fig:fig6]{Fig.~\ref*{fig:fig6}i}. 
	In other words, when the mirror nodes with the largest frequency differences in the right part are in-phase, their counterparts in the left part are in an anti-phase relation, and vice versa. The behavior of the counterpart mirror nodes increasingly resembles each other as their interlayer frequency differences decrease. This is particularly evident in the middle part, where the frequency difference between the mirror nodes in different layers is zero, resulting in identical behavior, as observed in \hyperref[fig:fig6]{Fig.~\ref*{fig:fig6}h}. The patterns observed in \hyperref[fig:fig6]{Fig.~\ref*{fig:fig6}h,i} do not appear in the intralayer couplings within the first hysteresis loop (see supplementary Fig.~SF5). To observe how the phase correlation between the $i^{th}$ pairs of mirror nodes $C_i(t)$ evolves over time and to understand the phase shift between the time series of counterpart mirror pairs  $C_i(t)$ and  $C_{N+1-i}(t)$ within each section, please refer to supplementary Fig.~SF6.

	The findings indicate that the superimposed waves and complex time-varying partial synchronization patterns observed in \hyperref[fig:fig6]{Fig.~\ref*{fig:fig6}a-g} arise from the information transfer between nodes in the right and left regions, both within and across each layer. It's important to note that while the mirror nodes may not be locked, the dynamics of the first layer are similar to those of the second layer~(see supplementary Fig.~SF7).

	\begin{figure}[!ht]
		\centering
		\includegraphics[width=0.82\linewidth]{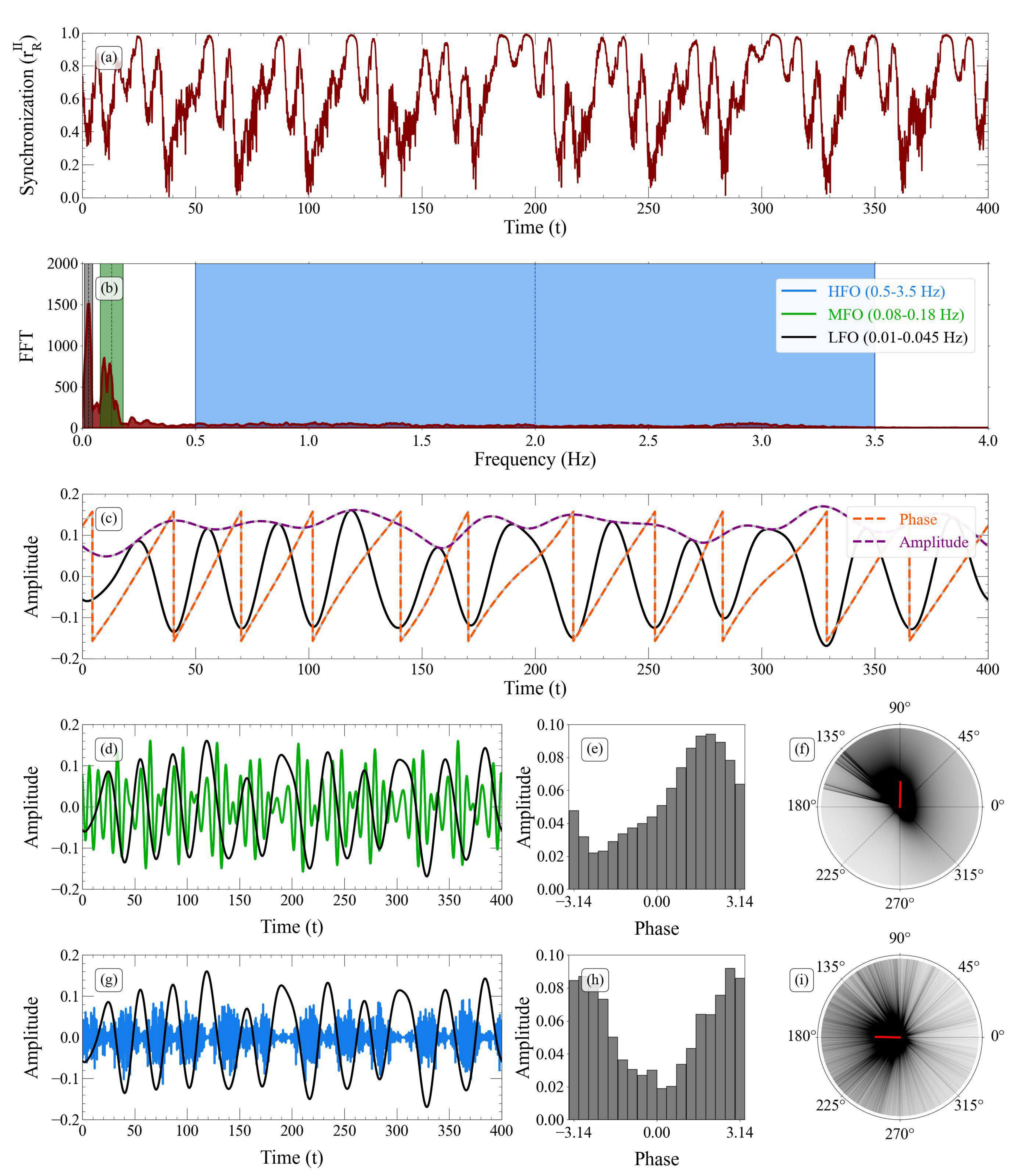}
		\caption{{Analysis of phase-amplitude coupling (PAC) across different frequency oscillations. ({\bf a}) The simulated signal displays the synchronization order parameter for the right part in layer~\Romannum{2} over time, utilizing parameters from \hyperref[fig:fig3]{Fig.~\ref*{fig:fig3}d}, with,  $\sigma=2.29$, within the second hysteresis loop. ({\bf b}) FFT visualization with color-coding for low (LFO), medium (MFO), and high (HFO) frequency oscillations. ({\bf c}) The black line represents the inverse Fourier transform of the LFOs, with the orange and purple dashed lines indicating its phase and amplitude. ({\bf d}-{\bf f}) PAC between LFOs and MFOs, measured using ({\bf e}) PLV and ({\bf f}) MVL modulation indices. ({\bf g}-{\bf i}) Comparable calculations for LFOs and HFOs.}}
		\label{fig:fig7}
	\end{figure}

	We noticed that the time evolution of the order parameter in each part exhibits various superimposed frequency waves. Our goal is to explore the potential coupling between phase and amplitude within these oscillations. To achieve this, we employed the Fast Fourier Transform (FFT) to decompose the signals into three distinct frequency bands: LFOs ranging from 0.01 to 0.045 Hz, MFOs between 0.08 and 0.18 Hz, and HFOs spanning 0.5 to 3.5 Hz (refer to  \hyperref[fig:fig7]{Fig.~\ref*{fig:fig7}a,b}). 
	Applying the Hilbert transform to each signal in various frequency bands allows us to find the phase and amplitude at each time point.
	\hyperref[fig:fig7]{Fig.~\ref*{fig:fig7}c} displays the LFOs extracted from the right part order parameter (black line), along with its phase (orange dashed line) and amplitude (purple dashed line). The fourth and fifth rows display the PAC between LFOs and MFOs, and between LFOs and HFOs, respectively. \hyperref[fig:fig7]{Fig.~\ref*{fig:fig7}d} shows the LFOs and MFOs signals and \hyperref[fig:fig7]{Fig.~\ref*{fig:fig7}g} shows the LFOs and HFOs signals. To visualize the interactions, the average amplitudes of MFOs and HFOs are plotted against the LFOs phase in \hyperref[fig:fig7]{Fig.~\ref*{fig:fig7}e,h}.

	\hyperref[fig:fig7]{Fig.~\ref*{fig:fig7}f,i}  illustrate phase clustering biases in polar space. Investigations confirm a clear connection between LFO phase and MFO and HFO amplitudes: MFO peaks at LFO phase $\frac{\pi}{2}$ and HFO peaks at $\pi$. The PLV for MFOs and HFOs is 0.315 and 0.307, respectively. Similarly, the MVL is 0.032 and 0.038, respectively. Comparable results are obtained using the signal from the right part.
	
	\begin{figure}[!ht]
		\centering
		\includegraphics[width=0.86\linewidth]{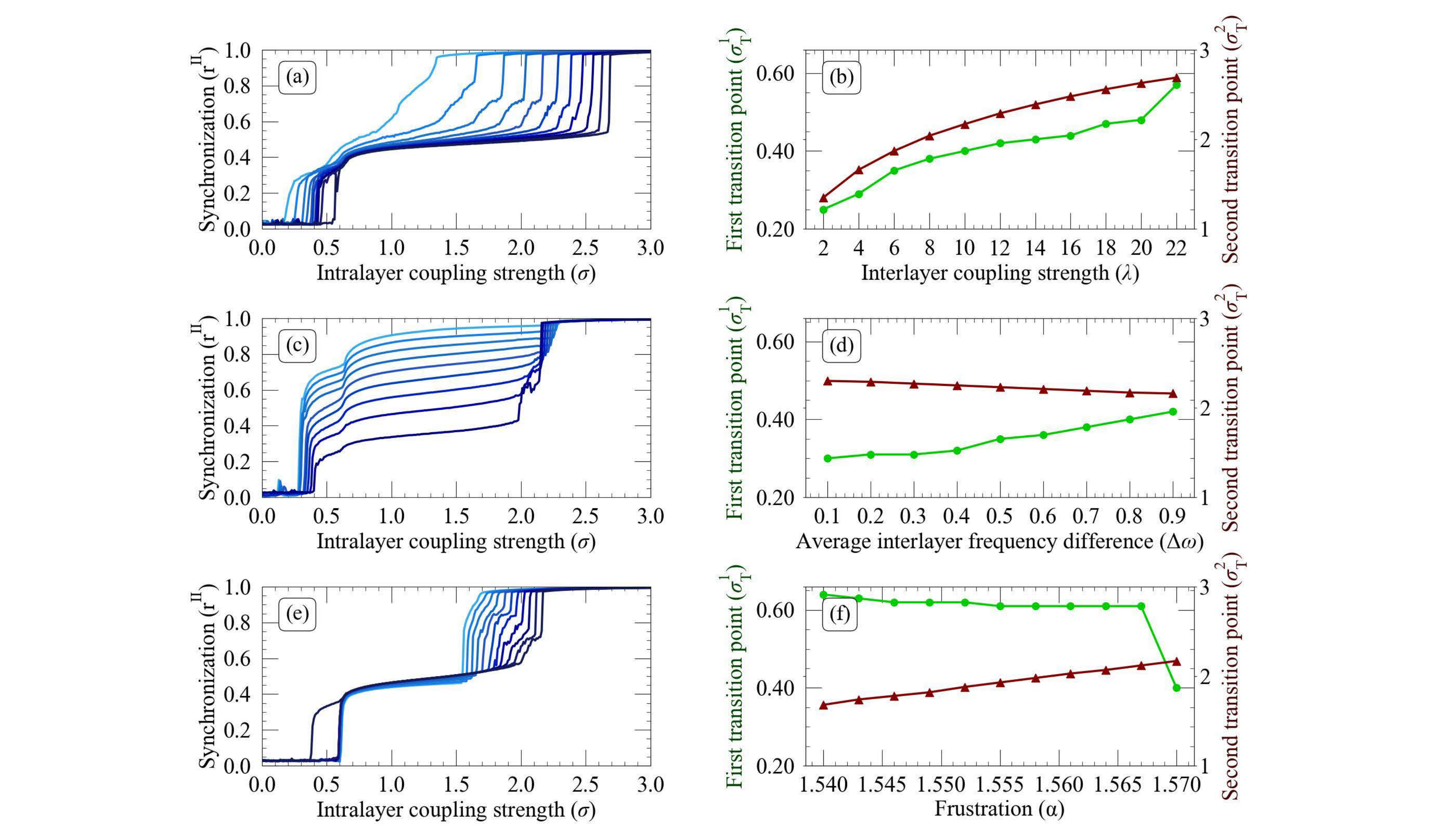}
		\caption{{Analysis of phase transition changes due to variations in model parameters. The left column depicts the synchronization phase transition in the backward path, while the right column shows critical intralayer coupling for the first (green circle) and second (red triangle) jumps. From top to bottom, each row examines the impact of modifying a single network parameter—interlayer coupling ($\lambda$), average natural frequencies of mirror nodes ($\Delta\omega$), and frustration ($\alpha$)—while keeping the others constant. The phase transition curves shift from light to dark blue, representing low to high values of the examined parameters. The parameters used are $\alpha= \frac{\pi}{2}$, $\lambda= 10$ and $\Delta\omega= 0.8$. }}
		\label{fig:fig8}
	\end{figure}

	So far, we have explored how the Regular arrangement of mirror node frequencies affects the dynamics of a duplex network with interlayer coupling set at $\lambda=10$, an average frequency difference of $\Delta\omega=0.8$, and interlayer frustration fixed at $\alpha = \frac{\pi}{2}$. Our findings showed two hysteresis loops in the transition curves, indicating complex dynamics. The rows of \hyperref[fig:fig8]{Fig.~\ref*{fig:fig8}} depict how these transition curves evolve in response to changes in the parameters $\lambda$, $\Delta\omega$, and $\alpha$, respectively. The first column shows the transition curve for different parameter values along the backward path. The blue shades, which range from light to dark, illustrate the transition diagrams for the parameter under study as it progresses from lower to higher levels. Meanwhile, the second column emphasizes the locations of the first jumps (green circles) and the second jumps (red triangles) in the transition curves in relation to the parameter values.

	The blue shades in \hyperref[fig:fig8]{Fig.~\ref*{fig:fig8}a}, ranging from light to dark, illustrate the transition diagrams for interlayer coupling levels as they progress from lower to higher values. In \hyperref[fig:fig8]{Fig.~\ref*{fig:fig8}b}, it is evident that with increasing interlayer coupling, the first and second transitions take place at higher intralayer coupling parameter values. As a result, the rise in the interlayer coupling coefficient postpones the transitions and leads to sharper transitions, emphasizing that the duplex structure of the network contributes to these atypical transitions.
	
	In \hyperref[fig:fig7]{Fig.~\ref*{fig:fig8}c,d}, we examine how average mirror node frequencies affect the phase transition. Small $\Delta\omega$ values create a more pronounced initial transition at lower intralayer coupling due to our model, which allocates more nodes to the middle section. A larger synchronized middle section leads to greater jumps during the first transition. Meanwhile, the smaller corner sections, which have a larger frequency difference, are less likely to disrupt the synchronization of the middle section. This facilitates synchronization in the middle section, resulting in a smaller transition point with a lower  $\Delta\omega$. Additionally, smaller values of $\Delta\omega$, resulting from the smaller sizes of the corner sections, lead to a decrease in the height of the second transition jumps associated with their synchronization.

	The earlier study indicated that frustration near $\frac{\pi}{2}$ leads to hysteresis loops~\cite{kumar2021explosive}. In \hyperref[fig:fig7]{Fig.~\ref*{fig:fig8}e,f}, we found that increasing frustration around $\frac{\pi}{2}$ decreases the first transition points while increasing the second ones. The effect of frustration change is more pronounced during the second transition, which involves the synchronization of the left and right sections. This is because these sections exhibit greater frequency differences between mirror nodes. In fact, as frustration approaches $\frac{\pi}{2}$, it becomes increasingly difficult for the left and right sections to synchronize with the entire network.

	\section*{Discussion}
	\phantomsection
	\label{Discussion}
	\addcontentsline{toc}{section}{Discussion}
	This study intricately investigates the dynamics of duplex networks, emphasizing the significance of frequency distributions among mirror nodes in shaping phase transitions and synchronization across network layers. Building upon earlier findings that underscored the potential for explosive synchronization due to frequency mismatches between network layers and the inclusion of $\frac{\pi}{2}$ frustration within interlayer connections in duplex networks~\cite{kumar2021explosive}, our study contributes novel insights to this domain.
	
	Expanding on the work by Kumar et al.~\cite{kumar2021explosive}, our research advances understanding by not only highlighting the pivotal role of average frequency differences between mirror nodes in phase transitions but also emphasizing the critical influence of the specific configuration of these frequencies on transition dynamics. We introduce a Regular frequency assignment model where each layer comprises consistent arrays of natural frequencies evenly distributed between  $-0.5$ and $0.5$. Nodes in each layer are classified into three groups: left, middle, and right, with mirror nodes in these groups displaying positive, neutral, and negative frequency differences  ($\delta\omega_i$, for $i=1,\dots, N$  ), respectively. By manipulating the sizes of these groups, we can control the average frequency differences and compare the behavior of this model with a Randomly frequency assignment model featuring equivalent average frequency discrepancies.
	
	Our preliminary findings are in line with the outcomes presented by Kumar et al.~\cite{kumar2021explosive}, demonstrating that random frequency differences among mirror nodes induce explosive synchronization when  $\alpha=\frac{\pi}{2}$. Conversely, this phenomenon leads to a continuous phase transition when $\alpha=0$.
	
	Our research uncovers that in the Regular frequency assignment model, a hysteresis loop emerges on the transition curve even when $\alpha=0$, indicating abrupt synchronization of nodes with $\delta\omega_i\neq0$ in the peripheral sections,  contrasting the gradual phase transition observed in the Random model at $\alpha=0$. This should be due to the multimodal distribution of frequency discrepancies.

	Previous studies have examined the impact of natural frequency distributions on single-layer networks. Martens et al. investigated globally connected networks with bimodal frequency distributions, uncovering states such as incoherence, partial synchrony, and standing waves with anti-phase groups, while highlighting memory and irreversibility in transition curves~\cite{martens2009exact}. In contrast, our study focuses on two globally connected layers with uniform frequency distributions. When merged into a duplex network using the Regular frequency assignment model, we observed irreversible traits in the phase transition curves that arise solely from the unique frequency allocation in the duplex structure.

	In the Regular frequency arrangement model with $\alpha=0$, abrupt synchronization occurs among peripheral nodes, distinguished by frequency differentials from their mirrored pairs, followed by reversible synchronization in the middle section, where node frequencies align precisely with their mirrored pairs. At $\alpha=\frac{\pi}{2}$, we observe double hysteresis loops. Unlike $\alpha=0$, at low intralayer couplings, the middle nodes are synchronized, while at higher couplings, the peripheral nodes achieve synchronization. In fact, frustration alters the synchronization order: with zero frustration, peripheral sections synchronize first, followed by the central section as intralayer coupling increases; this order is reversed at $\alpha=\frac{\pi}{2}$.  We should note that, unlike the Random frequency assignment model, in the Regular model, when $\alpha=\frac{\pi}{2}$, there is also an abrupt transition observed for nodes with zero frequency discrepancies from their mirrored pairs. An earlier study explored consecutive explosive transitions in multilayer networks, focusing on the interaction between a dynamical layer of phase oscillators and an approximately synchronized environmental layer. However, unlike our study, their analysis examined phase transitions by varying the interlayer coupling without considering the frequency discrepancy distribution.~\cite{wu2022double}.
	
	The detailed examination of the instantaneous correlation matrices for intralayer coupling across both hysteresis loops shows a periodic pattern in the synchronization of each layer. In the first hysteresis loop, at low intralayer couplings, this periodicity originates from the synchronization of the middle section.  In contrast, in the second hysteresis loop, the synchronization order parameter for the left and right sections behaves periodically.

	The periodic behavior observed in the initial hysteresis loops is straightforward and connected to the formation of synchronized groups with varying frequencies among the nodes in the middle section. In contrast, the periodic behavior observed in the second hysteresis loops—related to the synchronization of the left and right sections—displays a complex wave composed of multiple frequencies. This complexity arises from the intricate interactions between the left and right sections, both within and across layers.  Throughout this process, the synchronization of nodes in each left and right section shifts between coherent and incoherent states across different time scales, a phenomenon we refer to as the blinking process. Furthermore, our findings indicate that the overlapping waves are not independent; rather, the amplitudes of the higher frequency oscillatory components are selectively amplified at specific phases of the lower-frequency signal.
	
	In nature, many waveforms are complex and composed of multiple frequencies. Thus, exploring the structures that contribute to these synchronization patterns is intriguing. For instance, periodic signals in Electroencephalography (EEG) represent essential neural synchronization patterns vital for understanding brain states and cognitive functions. Additionally, these oscillations are not independent; they clearly exhibit signs of interaction~\cite{tort2010measuring,hanslmayr2014brain,voytek2010shifts}. For example, gamma activity has been shown to couple with opposing theta phases during both encoding and recall memory processes, as evidenced by local field potentials~\cite{saint2023gamma}.

	In conclusion, our study highlights the essential connection between frequency distribution and network dynamics in two-layer systems. Given that many real-world networks, including those in the brain, are multilayered and consist of heterogeneous entities with varying firing frequencies, our findings provide a vital framework for further exploration of these complex systems. For example, various complex behaviors observed in networks, such as phase-amplitude coupling, may arise from the interplay between frequency distribution, phase lag, and multilayer interactions. We emphasize the necessity for ongoing research in this area to deepen our understanding of these intricate dynamics.
	
	\section*{Data availability}
	\phantomsection
	\label{Data availability}
	\addcontentsline{toc}{section}{Data availability}
	The datasets generated and analysed during the current study are available in the Github repository, \href{https://github.com/Articles-data/Frequency-Arrangement}{https://github.com/Articles-data/Frequency-Arrangement}.

	~

	\section*{Author contributions}
	\phantomsection
	\label{Author contributions}
	\addcontentsline{toc}{section}{Author contributions}
	M.Z. designed the study and developed the initial concept. A.S. conducted the simulations, performed the analysis, and prepared the results. Both authors discussed the results and contributed to writing the manuscript.

	\section*{Competing interests}
	\phantomsection
	\label{Competing interests}
	\addcontentsline{toc}{section}{Competing interests}
	The authors declare no competing interests.
	
	\section*{Additional information}
	\phantomsection
	\label{Additional information}
	\addcontentsline{toc}{section}{Additional information}
	
	%{\bf Supplementary Information} The online version contains supplementary material available at \textcolor{green}{https://doi.\allowbreak org/}.
	%\newline
	{\bf Correspondence} and requests for materials should be addressed to M.Z.

	\FloatBarrier

\newpage

% Redefine \biblabel to make numbers bold in the bibliography
\makeatletter
\renewcommand\@biblabel[1]{\textbf{#1.}\hfill}
\makeatother

	\maketitle %Generates the title block using the current formatting. 
	\pagenumbering{arabic} %Switch to Arabic numerals
	
	\setcounter{page}{1} % Reset page counter for supplementary
	\renewcommand{\thepage}{S\arabic{page}} % Change page numbering to S1, S2, S3, ...

	\newpage
	\section*   {Supplementary materials for double hysteresis loop in synchronization transitions of multiplex networks: the role of frequency arrangements and frustration}
	
		This supplementary material provides explanations for supplementary figures SF1–SF7 and supplementary videos SV1–SV3.
	\section*{Description of Supplementary Figures }
	\phantomsection
	\label{Description of Supplementary Figures}
	\addcontentsline{toc}{section}{Description of Supplementary Figures}
	\subsection*{Description of Figure SF1:}
	\phantomsection
	\label{Figure SF1}
	\addcontentsline{toc}{subsection}{Figure SF1}
	This figure illustrates that when there is no phase lag ($\alpha=0$), the observed jump during the phase transition is linked to the peripheral sections, where nodes display frequency differences compared to their corresponding mirror nodes in the opposite layer. The Regular frequency assignment model was utilized in this analysis. The synchronization phase transition is represented in both forward and backward paths for the entire layer~\Romannum{2}~(blue), as well as for the left~(green), middle~(black), and right~(red) sections. As depicted, the transition is continuous and reversible for nodes in the middle section, while the left and right sections exhibit discontinuous phase transitions with hysteresis loops.

	\begin{figure}[!ht]
		\renewcommand{\figurename}{Figure SF}
		\centering
		\includegraphics[width=0.9\linewidth]{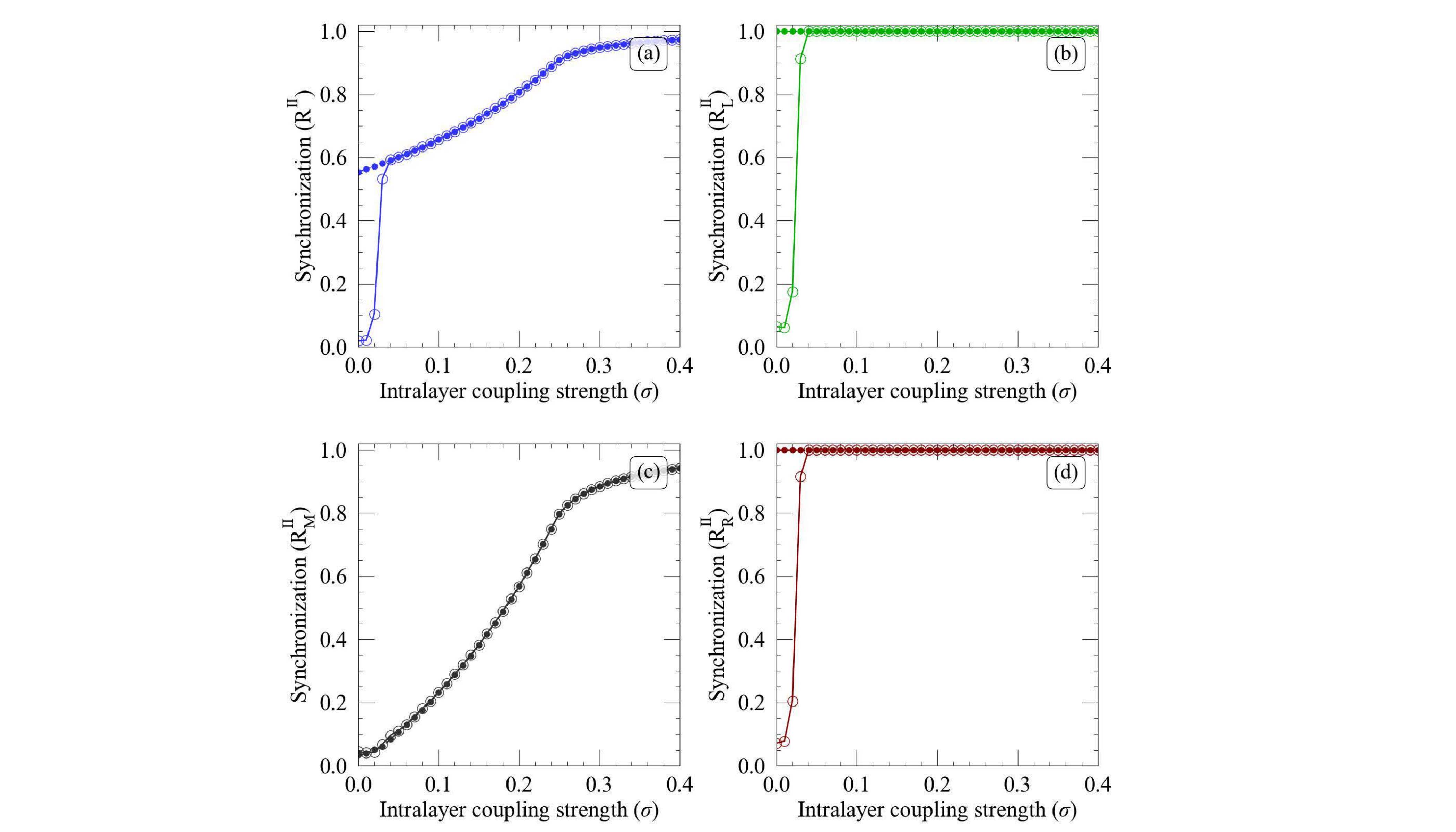}
		\caption{Investigation of phase transitions from incoherent to coherent states across the entire layer and sections of layer~\Romannum{2} using a Regular frequency arrangement model with $\alpha=0$. Panel ({\bf a}) shows the transition curve for the entire layer, while panels ({\bf b}), ({\bf c}), and ({\bf d}) depict the transition curves for the left, middle, and right sections, respectively. Forward synchronization is indicated by solid lines with open markers, and backward synchronization by dashed lines with filled markers. Parameters are  $\Delta \omega=0.8$ and $\lambda=10$.}
		\label{fig:FigureSF1}
	\end{figure}
	
	\newpage
	\subsection*{Description of Figure SF2:}
	\phantomsection
	\label{Figure SF2}
	\addcontentsline{toc}{subsection}{Figure SF2}
	The purpose of this figure is to demonstrate that in the absence of a phase lag ($\alpha=0$), the peripheral sections remain synchronized during the backward transition, even with zero intralayer coupling strength ($\sigma=0$), due to interlayer interactions. Panels (a) to (d) display the instantaneous correlation matrices ($D_{i j}(t)= \cos \left(\theta_i(t)-\theta_j(t)\right)$) for layer~\Romannum{2} along the backward paths, corresponding to  $\sigma=0.5$,  $\sigma=0.25$, $\sigma=0.15$, and  $\sigma=0$,  respectively. At high values of intralayer coupling, the network approaches synchronization. As $\sigma$ decreases, the middle section—where nodes have zero frequency differences with their mirror nodes—loses synchrony, while the left and right sections remain synchronized even at $\sigma=0$. Regular frequency assignment model was employed in this analysis.

	\begin{figure}[!ht]
		\renewcommand{\figurename}{Figure SF}
		\centering
		\includegraphics[width=0.9\linewidth]{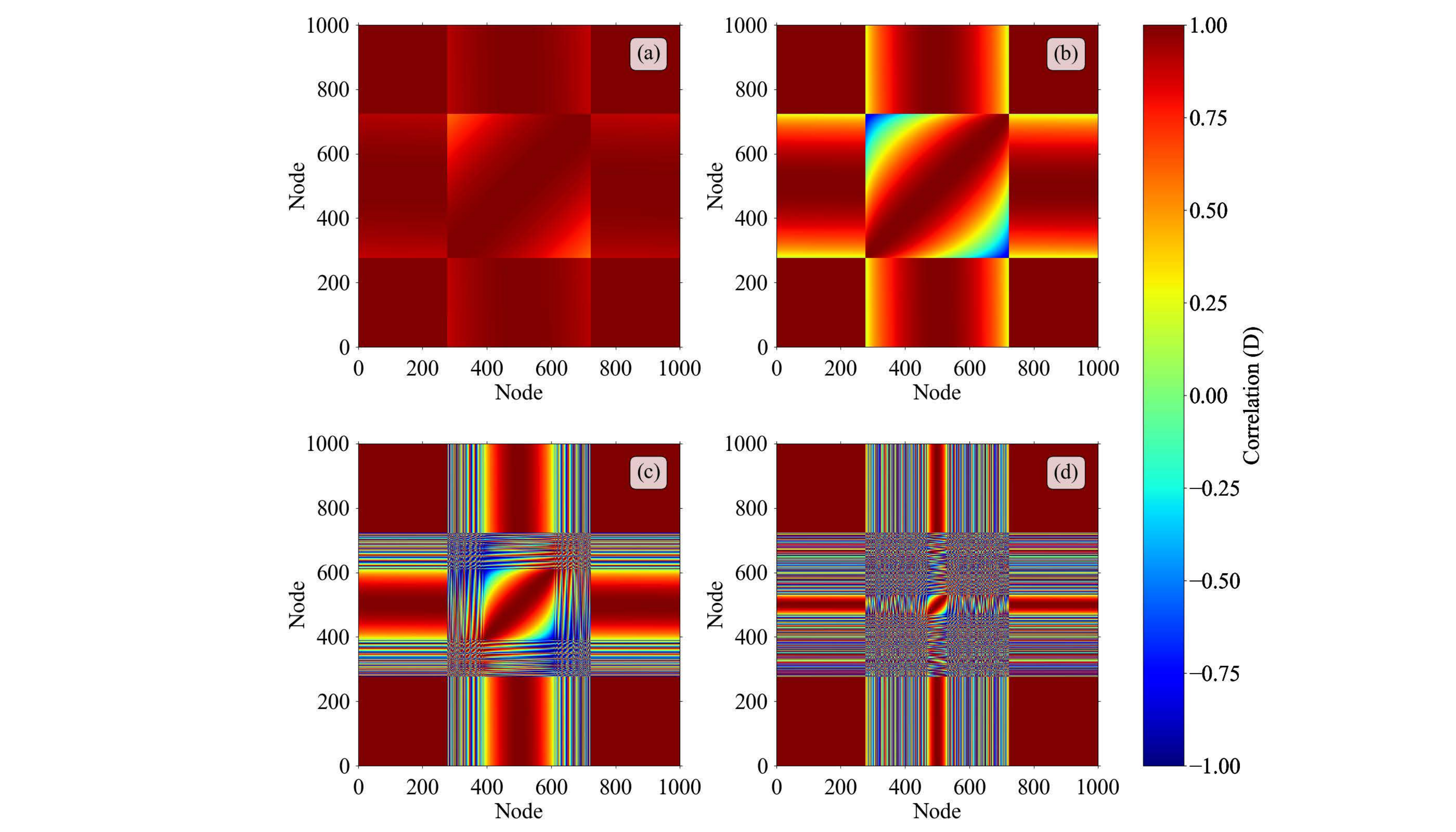}
		\caption{ Instantaneous correlation matrices for layer~\Romannum{2} across varying intralayer coupling strengths in the backward paths, with $\alpha=0$. Panels ({\bf a}) to ({\bf d}) represent ({\bf a}) $\sigma=0.5$, ({\bf b}) $\sigma=0.25$, ({\bf c}) $\sigma=0.15$, and ({\bf d}) $\sigma=0$. Parameters are  $\Delta \omega=0.8$ and $\lambda=10$.}
		
		\label{fig:FigureSF2}
	\end{figure}
	\newpage
	
	\subsection*{Description of Figure SF3:}
	\phantomsection
	\label{Figure SF3}
	\addcontentsline{toc}{subsection}{Figure SF3}
	This figure illustrates that at $\alpha=\frac{\pi}{2}$, the initial jump observed during the phase transition corresponds to the synchronization of the middle section, where nodes have zero frequency differences compared to their mirror nodes in the opposite layer. The subsequent hysteresis loop reflects the synchronization of the left and right sections, where nodes exhibit frequency differences relative to their mirrored ones. The Regular model for frequency assignments was employed in this analysis. The synchronization phase transition is presented for both forward and backward paths for the entire layer (blue), as well as for the left (green), middle (black), and right (red) sections. This transition is characterized as discontinuous and irreversible for nodes in both the middle and peripheral sections. In fact, during the first hysteresis loop, the middle section oscillates between synchronized and asynchronous states. As the intralayer couplings increase, just prior to the onset of the second hysteresis loop, the middle section achieves full synchronization in a reversible manner.

	\begin{figure}[!ht]
		\renewcommand{\figurename}{Figure SF}
		\centering
		\includegraphics[width=0.9\linewidth]{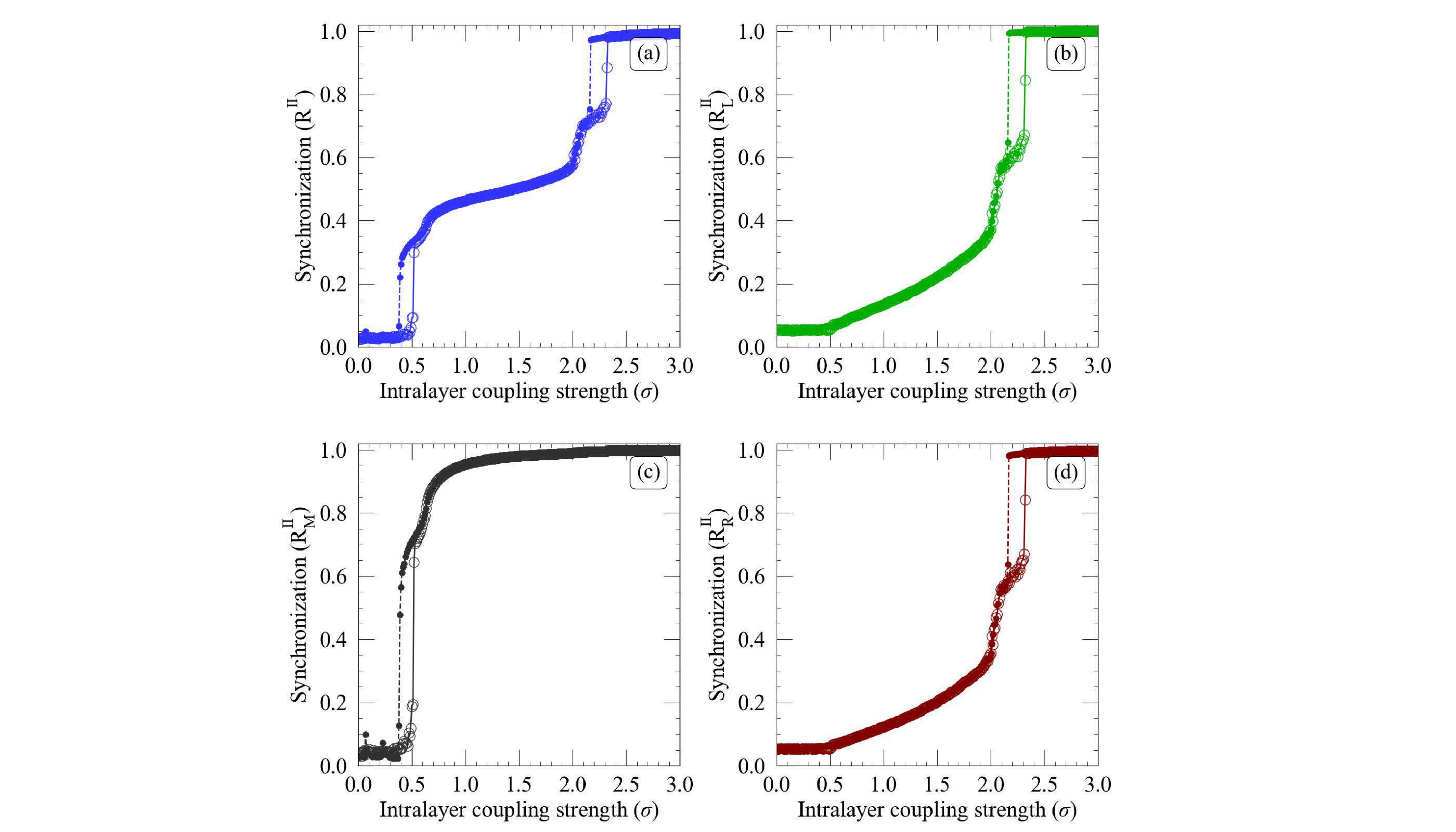}
		\caption{
			Investigation of phase transitions from incoherent to coherent states across the entire layer and sections of layer~\Romannum{2} using a Regular frequency arrangement model with $\alpha=\frac{\pi}{2}$. Panel ({\bf a}) shows the transition curve for the entire layer, while panels ({\bf b}), ({\bf c}), and ({\bf d}) depict the transition curves for the left, middle, and right sections, respectively. Forward synchronization is indicated by solid lines with open markers, and backward synchronization by dashed lines with filled markers. Parameters are  $\Delta \omega=0.8$ and $\lambda=10$.}
		\label{fig:FigureSF3}
	\end{figure}

	\newpage
	\subsection*{Description of Figure SF4:}
	\phantomsection
	\label{Figure SF4}
	\addcontentsline{toc}{subsection}{Figure SF4}
	This figure illustrates that at the stationary state, all pairs of mirror nodes in the middle section are synchronized, while the phase difference between mirror nodes in the left and right sections varies periodically. This periodic variation is particularly pronounced when $\alpha=\frac{\pi}{2}$. The figure illustrates the phase correlations for all pairs of mirror nodes, defined as $C_i(t)=\cos(\theta_i^{I}(t)-\theta_i^{II}(t))$, for $i=1\dots N$. Panel (a) shows results for  $\sigma = 0.39$, while panel (b) presents results for  $\sigma = 2.11$. The narrow red strip in the center of both panels represents the nodes in the middle section that are synchronized with their corresponding mirror nodes. The Regular frequency assignment model was used in this analysis.
	
	\begin{figure}[!ht]
		\renewcommand{\figurename}{Figure SF}
		\centering
		\includegraphics[width=0.9\linewidth]{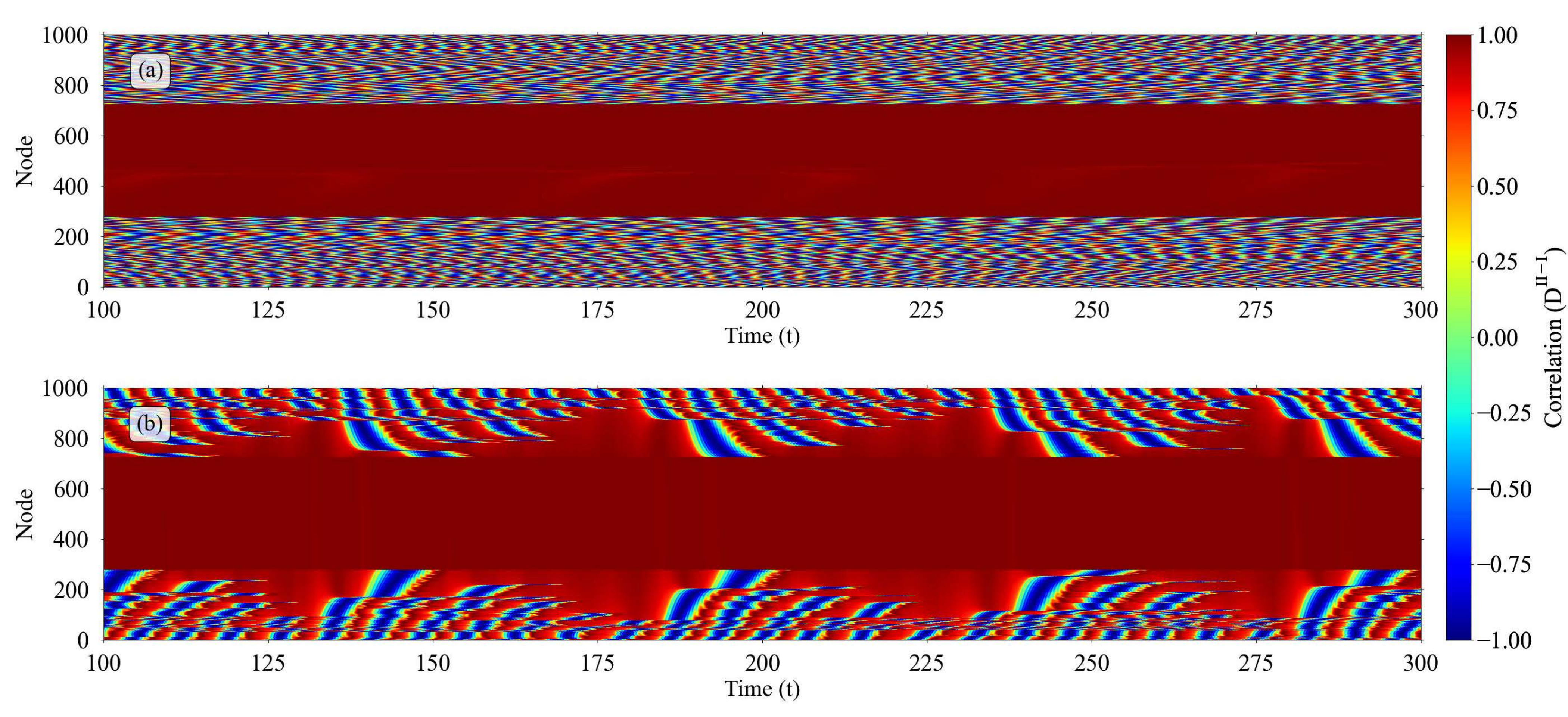}
		\caption{Phase correlation of mirror nodes over time, $C_i(t)=\cos(\theta_i^{I}(t)-\theta_i^{II}(t))$, for a system with $\alpha=\frac{\pi}{2}$. Panel ({\bf a}) displays the correlations for coupling within the first hysteresis loop at $\sigma = 0.39$, while panel ({\bf b}) shows them for the second hysteresis loop at $\sigma = 2.11$. Parameters are  $\Delta\omega=0.8$ and $\lambda=10$.}
		\label{fig:FigureSF4}
	\end{figure}

	\newpage
	\subsection*{Description of Figure SF5:}
	\phantomsection
	\label{Figure SF5}
	\addcontentsline{toc}{subsection}{Figure SF5}
	
	This figure illustrates that when the synchrony signals of the left and right sections (panel a) are separated by frequency, the low-frequency oscillations (LFOs) of both sections are in-phase (panel b), whereas the medium-frequency oscillations (MFOs) are in anti-phase (panel c). This analysis employed for a Regular model frequency assignment.
	
	\begin{figure}[!ht]
		\renewcommand{\figurename}{Figure SF}
		\centering
		\includegraphics[width=0.9\linewidth]{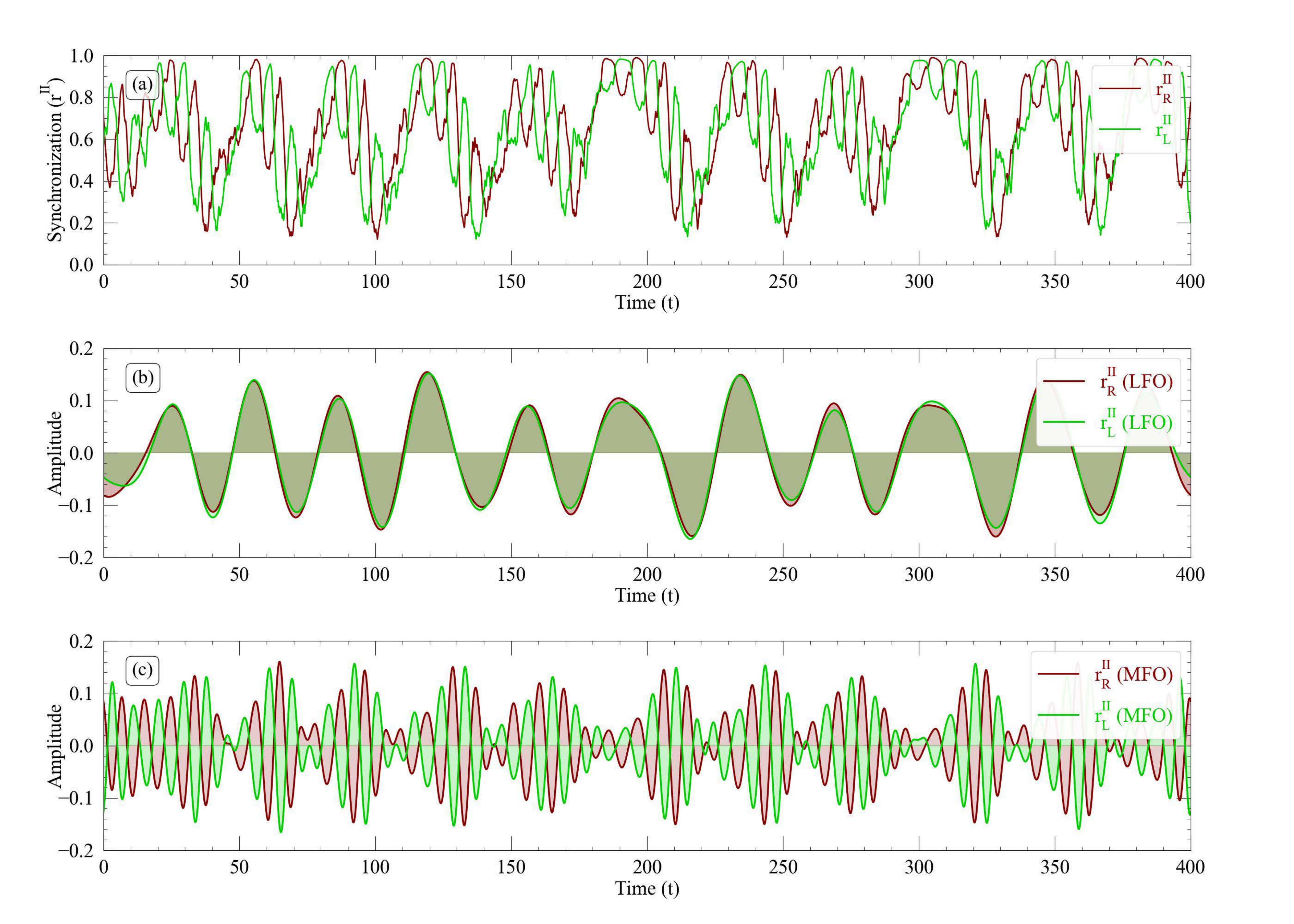}
		\caption{Phase correlations of synchronization order parameters in the left and right sections of layer~\Romannum{2}. ({\bf a}) Time evolution of local order parameters for the left (green) and right (red) sections. Frequency-filtered signals: ({\bf b}) low-frequency oscillations (LFOs) in the range of 0.01–0.045 Hz and ({\bf c}) middle-frequency oscillations (MFOs) in the range of 0.08–0.18 Hz.  The parameters used are $\Delta\omega=0.8$, $\lambda=10$, and $\sigma = 2.11$.}
		\label{fig:FigureSF5}
	\end{figure}
	
	\newpage
	
	\subsection*{Description of Figure SF6:}
	
	\phantomsection
	\label{Figure SF6}
	\addcontentsline{toc}{subsection}{Figure SF6}
	Each panel shows the phase correlations of two pairs of counterpart mirror nodes, which oscillate with the same frequency. However, as the magnitude of the frequency discrepancy between the mirror nodes ($|\delta\omega|$) decreases, the frequency of their correlation curves also decreases, until there is no oscillation when $|\delta\omega| = 0$ in panel d. Additionally, the two curves corresponding to the counterpart pairs become in-phase as the frequency discrepancy between the mirror nodes reduces (see panels a to d for illustration). 
	
	Thus, counterpart pairs located in the middle section are synchronized (panel d). However, when these counterpart pairs are positioned in different sections (left and right), the correlation oscillations of their corresponding mirror nodes are out of phase (panels a-c). When $|\delta\omega|$ is at its maximum, the two counterpart correlation curves become anti-phase~(panel a). The Regular frequency assignment model was employed in this analysis.

	\begin{figure}[!ht]
		\renewcommand{\figurename}{Figure SF}
		\centering
		\includegraphics[width=0.9\linewidth]{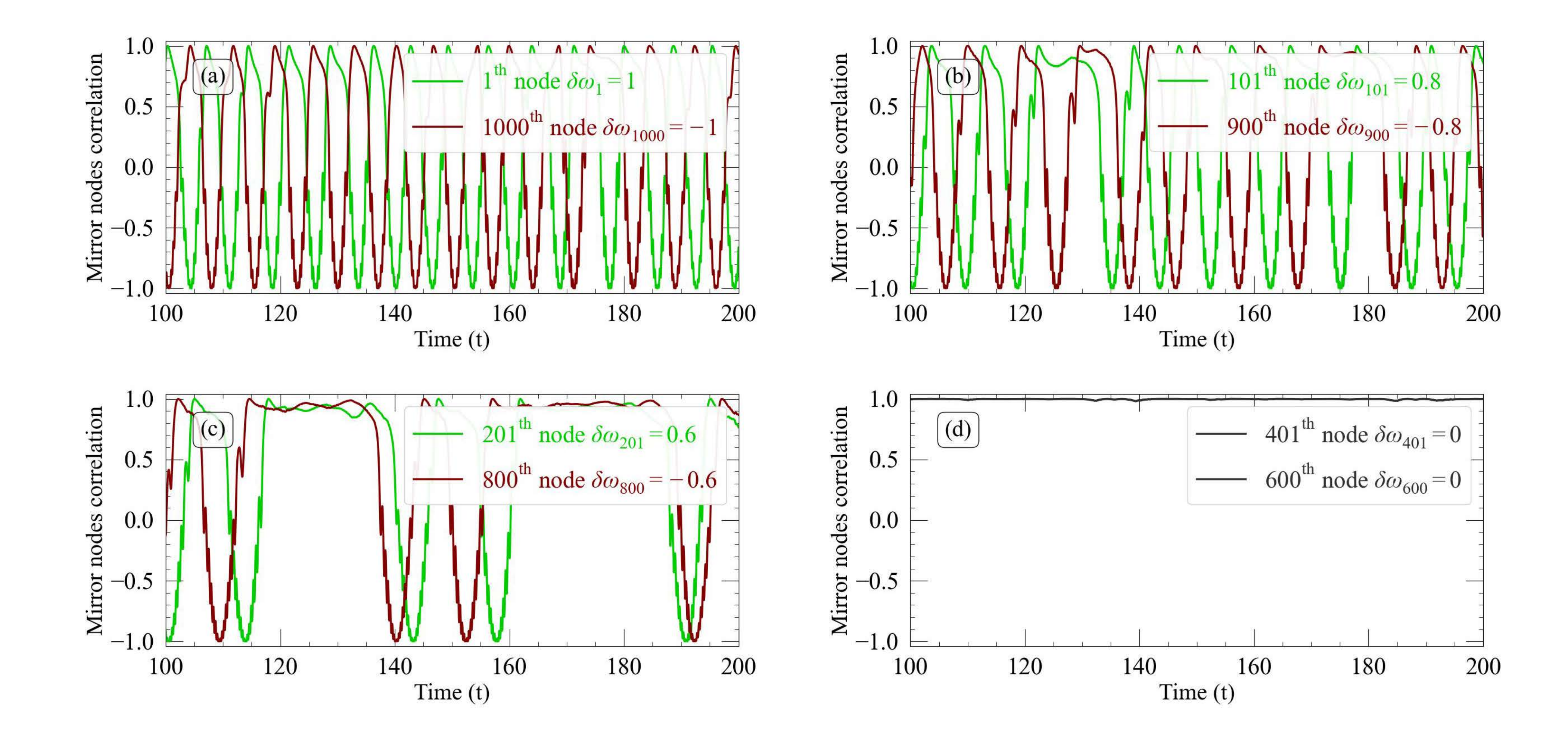}
		\caption{	
			Comparison of phase correlations ($C_i(t)=\cos(\theta_i^{I}(t)-\theta_i^{II}(t))$, where  $i$ labels  mirror pairs) for four distinct pairs of counterparts. 		
			Panel ({\bf a}) shows the correlations for pairs $i=1$ and $i=1000$; panel ({\bf b}) for pairs $i=101$ and $i=900$; panel ({\bf c}) for pairs $i=201$ and $i=800$; and panel ({\bf d}) repeats the correlation for pairs $i=401$ and $i=600$. The colors green, red, and black represent the correlations for mirror nodes in the left, right, and middle sections of the duplex network, respectively. The parameters used are $\Delta\omega=0.8$, $\lambda=10$, and $\sigma = 2.11$.}
		\label{fig:FigureSF6}
	\end{figure}
	\newpage

	\subsection*{Description of Figure SF7:}
	\phantomsection
	\label{Figure SF7}
	\addcontentsline{toc}{subsection}{Figure SF7}
	In our paper, we presented the results for the  layer~\Romannum{2}. This figure demonstrates that partial and complete synchronization in both layers are qualitatively similar. To this end, we have plotted the synchronization order parameter for the left, middle, and right sections, as well as for the entire layer. Panel (a) shows the results for layer~\Romannum{1}, and panel (b) for layer~\Romannum{2}. The Regular model for frequency assignments was used in this analysis.
	
	\begin{figure}[!ht]
		\renewcommand{\figurename}{Figure SF}
		\centering
		\includegraphics[width=0.9\linewidth]{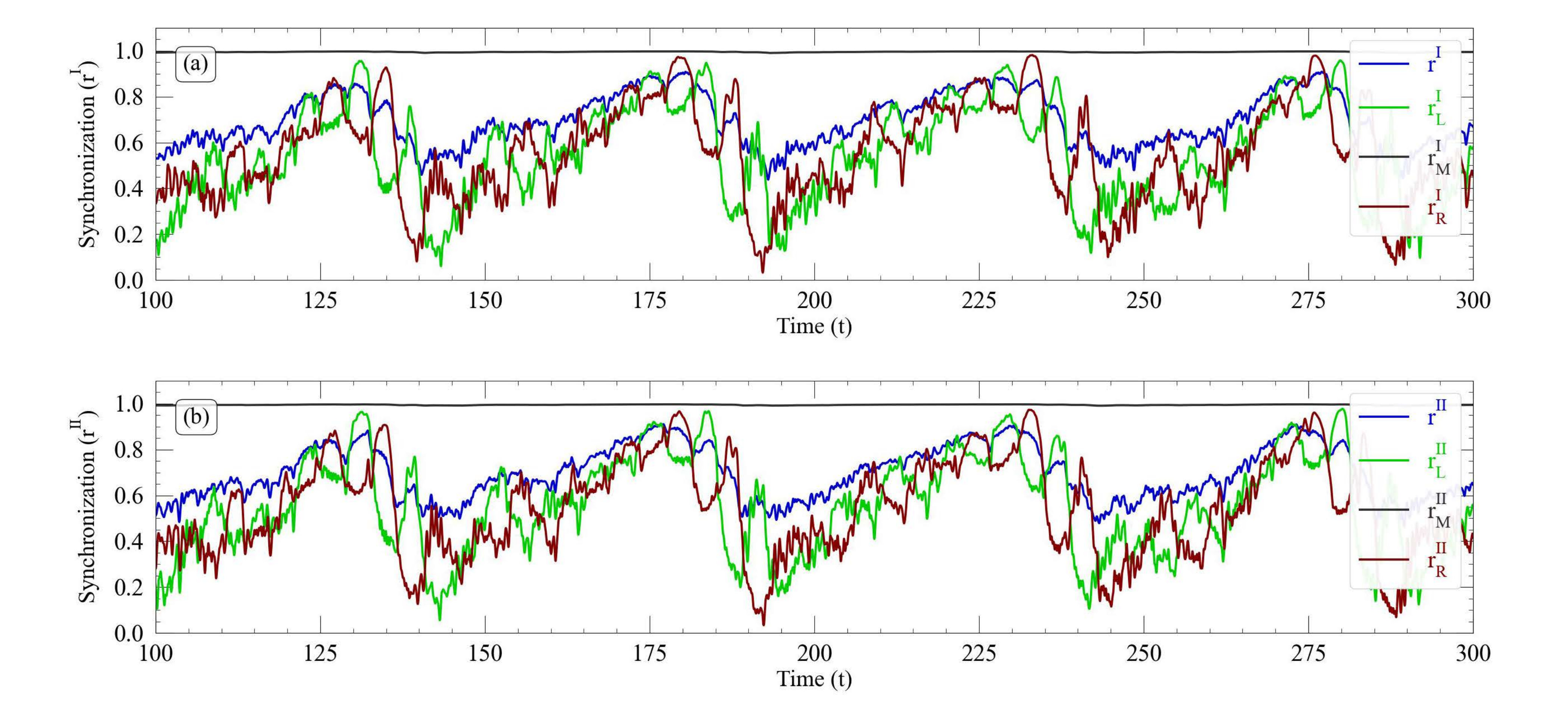}
		\caption{Comparison of the evolution of local and global synchronization order parameters for ({\bf a}) layer~\Romannum{1} and ({\bf b}) layer~\Romannum{2}. The blue line represents the synchronization of the entire layer, while the green, black, and red lines illustrate synchronization in the left, middle, and right sections of each layer, respectively. The parameters used are $\Delta\omega=0.8$, $\lambda=10$, and $\sigma = 2.11$.}
		\label{fig:FigureSF7}
	\end{figure}
	
	\newpage
	
	\section*{Description of Supplementary Videos }	
	\phantomsection
	\label{Description of Supplementary Videos}
	\addcontentsline{toc}{section}{Description of Supplementary Videos}
	In this section, we have taken snapshots from the supplementary videos to better illustrate and explain the panels.
	\newpage
	\subsection*{Description of video SV1: }
	\phantomsection
	\label{Figure SV1}
	\addcontentsline{toc}{subsection}{Figure SV1}
	The purpose of this video is to demonstrate the stationary dynamics of the duplex network with Regular frequency assignment, where $\sigma = 0$ and $\alpha = 0$. As we follow the backward paths, we observe that even when the intralayer coupling is zero, the right and left sections remain synchronized. This behavior is associated with the hysteresis loops observed in the absence of frustration.
	\begin{figure}[!ht]
		\renewcommand{\figurename}{Figure SV} % Change "Figure" to "Figure SV"
		\setcounter{figure}{0} % Reset the counter for figures
		\centering
		\includegraphics[width=0.9\linewidth]{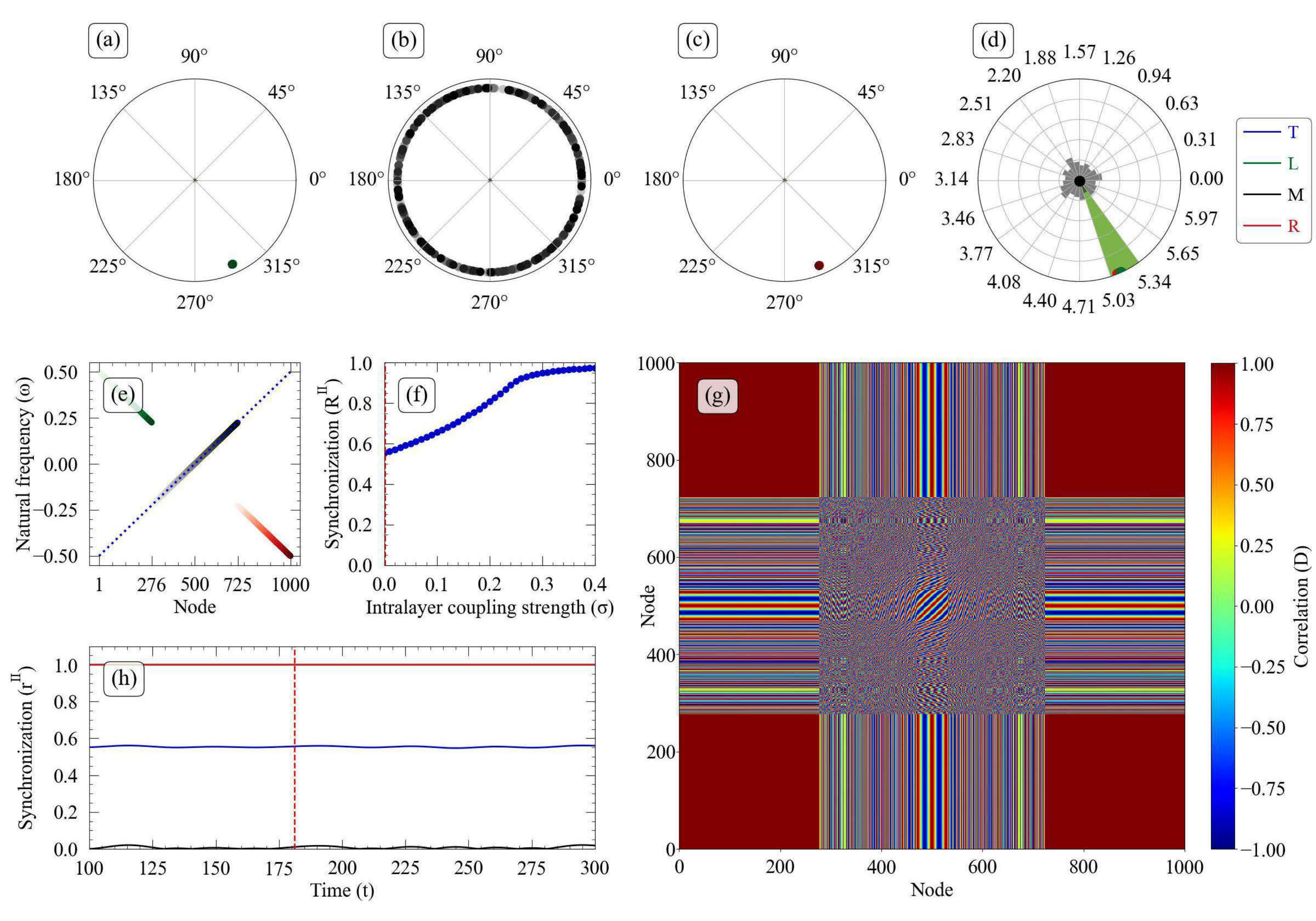}
		\caption{Stationary dynamics of layer~\Romannum{2} in a duplex network with zero intralayer coupling ($\sigma = 0$) and no frustration ($\alpha = 0$) under the Regular frequency assignment model. (First row): Panels~({\bf a–c}) show the phase evolution of nodes on the unit circle for the left (green points), middle (black points), and right (red points) sections, with the color of the nodes becoming more intense as the node index increases within each group. Panel~({\bf d}) displays the histogram of the nodes' phases in each group on the unit circle, with histograms for the left, middle, and right sections represented by green, black, and red, respectively. represented by its corresponding color. Panel~({\bf e}) shows the Regular frequency arrangement of nodes in the duplex network. The blue dashed line represents the natural frequencies of the nodes in layer~\Romannum{1}, with the nodes indexed in ascending order of their frequencies. The nodes in layer~\Romannum{2} are grouped into three distinct sets, indicated by green, black, and red, corresponding to their respective frequency values. Panel~({\bf f}) illustrates the phase transition from coherent to incoherent states along the backward path, with the red dashed line marking the intralayer coupling strength at the point of interest. Panel~({\bf g}) shows the time evolution of the correlation matrix ($D_{ij}(t) = \cos(\theta_i(t) - \theta_j(t))$) at the stationary state. Panel~({\bf h}) displays the time evolution of the synchronization order parameter for the left (green), middle (black), right (red), and entire layer (blue). The red dashed vertical line  marks the time of the dynamics in the videos. All the plots presented here are plotted along the backward paths with $\Delta\omega = 0.8$.}
		\label{fig:figSV1supp}
	\end{figure}
	\newpage	
	\subsection*{Description of video SV2: }
	\phantomsection
	\label{Figure SV2}
	\addcontentsline{toc}{subsection}{Figure SV2}
	The purpose of this video is to showcase the periodic behavior observed in the stationary dynamics of the duplex network with Regular frequency assignment, within the first hysteresis loop, where $\sigma = 0.39$ and $\alpha =\frac{\pi}{2}$. It is evident that the periodic behavior is primarily driven by the dynamics of the middle section.
	\begin{figure}[!ht]
		\renewcommand{\figurename}{Figure SV}
		\centering
		\includegraphics[width=0.9\linewidth]{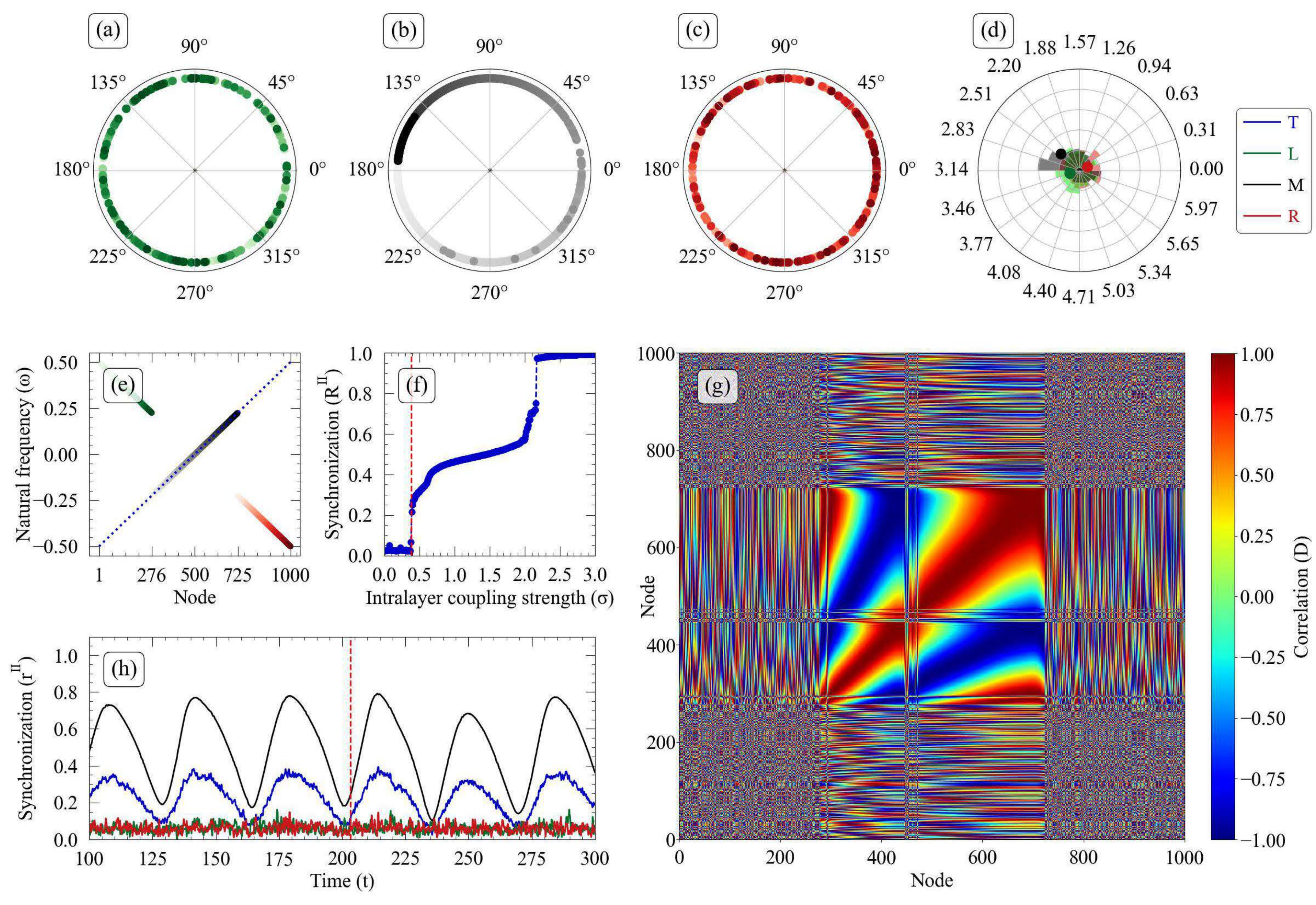}
		\caption{Stationary dynamics of layer~\Romannum{2} in a duplex network within the first hysteresis loop ($\sigma = 0.39$, $\alpha = \frac{\pi}{2}$) under the Regular frequency assignment model. The panel details are the same as those in supplementary video SV1.}
		\label{fig:figSV2supp}
	\end{figure}
	\newpage

	\subsection*{Description of video SV3: }
	\phantomsection
	\label{Figure SV3}
	\addcontentsline{toc}{subsection}{Figure SV3}
	The purpose of this video is to demonstrate the periodic behavior observed in the stationary dynamics of the duplex network with Regular frequency assignment, within the second hysteresis loop, where $\sigma = 2.11$ and $\alpha = \frac{\pi}{2}$. The blinking process is evident in the dynamics of the left and right sections, creating a complex wave with different frequencies in the synchronization order parameter of each section. 
	\begin{figure}[!ht]
		\renewcommand{\figurename}{Figure SV}
		\centering
		\includegraphics[width=0.9\linewidth]{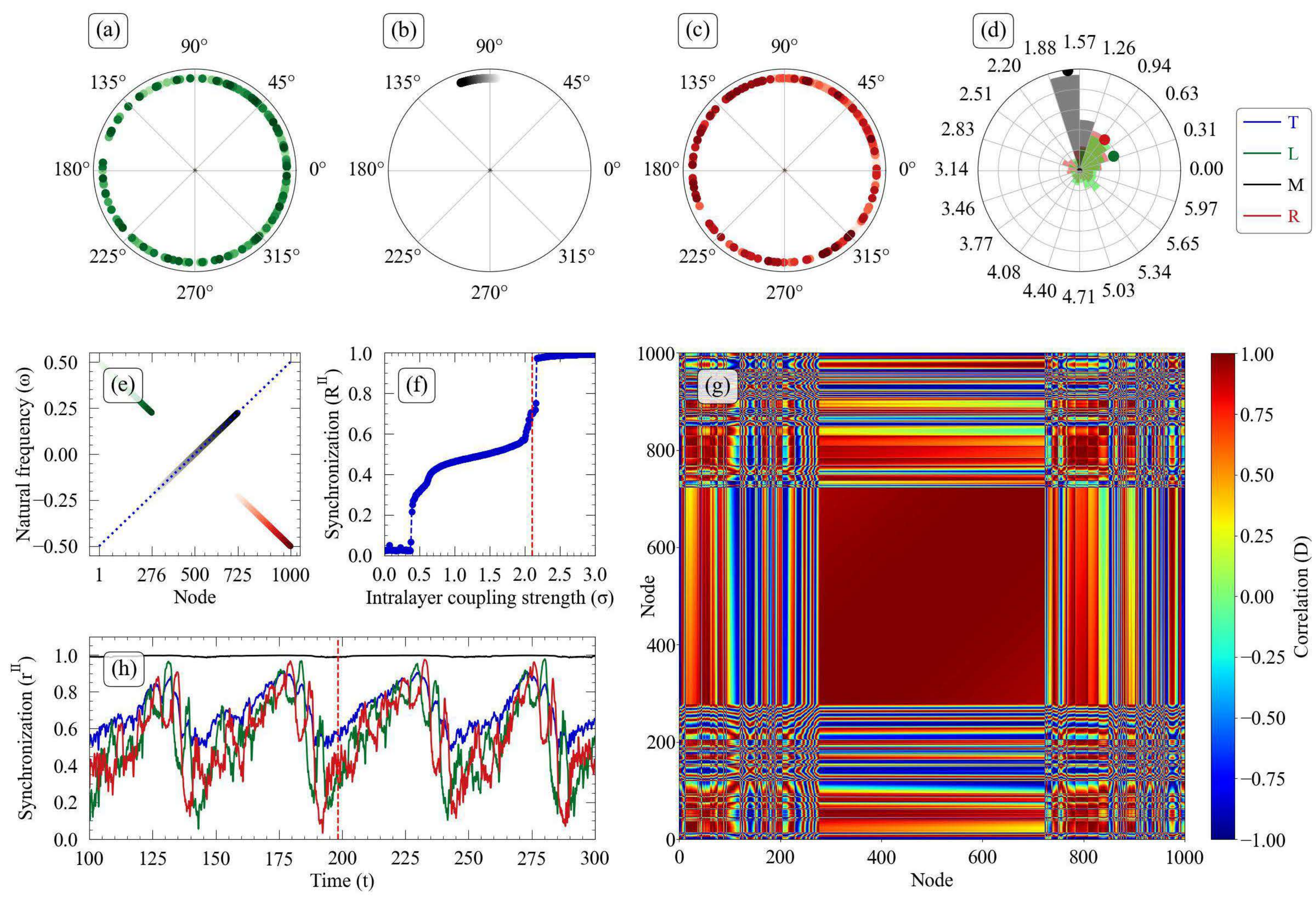}
		\caption{Stationary dynamics of layer~\Romannum{2} in a duplex network within the second hysteresis loop ($\sigma = 2.11$, $\alpha = \frac{\pi}{2}$) under the Regular frequency assignment model. The details of the panels are identical to those in supplementary video SV.1.}
		\label{fig:figSV3supp}
	\end{figure}
	\newpage

\end{document}